\documentclass[a4paper,11pt,notitlepage]{article}
\usepackage{a4wide,epsfig,graphics,amsthm,amsfonts,amssymb,amsopn,amsmath,longtable,dcolumn,calc,verbatim,threeparttable,eurosym,lscape,rotating,multirow, color, ulem, threeparttablex, environ,booktabs,verbatim,eurosym,pdflscape,rotating,wasysym,threeparttablex,bbm}
\usepackage[paper=a4paper,left=1in,right=1in,top=1in,bottom=1in]{geometry}
\usepackage[latin1]{inputenc}
\usepackage{lscape}
\usepackage{natbib}
\usepackage{ulem}
\usepackage{caption}
\usepackage{ragged2e,array,longtable}
\usepackage{arydshln}
\usepackage[shortlabels]{enumitem}

\usepackage[onehalfspacing]{setspace}
\newcolumntype{.}{D{.}{.}{-1}}
\newcolumntype{d}[1]{D{.}{.}{#1}}

\renewcommand{\baselinestretch}{1.5} 
\usepackage[colorlinks=true,linkcolor=blue,citecolor=blue,backref=page]{hyperref}
\usepackage{graphicx}
\usepackage{subfig}
\usepackage{changepage}
\usepackage{titletoc}

\newcommand{\pa}{\partial}

\newcommand{\bi}{\begin{itemize}}
\newcommand{\ei}{\end{itemize}}
\newcommand{\mc}{\multicolumn}
\newcommand{\bc}{\begin{center}}
\newcommand{\ec}{\end{center}}

\newcommand{\bs}{\begin{scriptsize}}
\newcommand{\es}{\end{scriptsize}}
\newcommand{\beq}{\begin{equation}}
\newcommand{\eeq}{\end{equation}}
\newcommand{\ben}{\begin{enumerate}}
\newcommand{\een}{\end{enumerate}}
\newcommand{\bean}{\begin{eqnarray}}
\newcommand{\eean}{\end{eqnarray}}

\DeclareMathSymbol{\R}{\mathalpha}{AMSb}{"52}
\DeclareMathSymbol{\E}{\mathalpha}{AMSb}{"45}
\newcommand{\sym}[1]{$^{#1}$} % for significance stars in Table

\usepackage{amssymb}
\usepackage{pifont}
\usepackage{float}

\def\sym#1{\ifmmode^{#1}\else\(^{#1}\)\fi}

\usepackage[section]{placeins}
\newlength{\asdf}
\usepackage{xargs}   % Use more than one optional parameter in a new commands
\usepackage[pdftex,dvipsnames]{xcolor}  % Coloured text etc.
\usepackage[colorinlistoftodos,prependcaption,textsize=tiny]{todonotes}
\newcommandx{\MC}[2][1=]{\todo[inline, linecolor=blue,backgroundcolor=blue!25,bordercolor=blue,#1]{#2}} % Marco
\newcommandx{\RM}[2][1=]{\todo[inline, linecolor=Plum,backgroundcolor=Plum!25,bordercolor=Plum,#1]{#2}} % RM
\newcommand{\AS}[1]{{\color{Orange}{#1}}}

\begin{document}
\startcontents[sections]
%\printcontents[sections]{l}{1}{\setcounter{tocdepth}{2}}

\renewcommand{\baselinestretch}{1.12}
\title{\vspace{-10mm}\textbf{The Accuracy of Job Seekers' Wage Expectations}\thanks{We thank Patrick Arni, Adeline Delavande, Ingo Isphording, Jonas Jessen, Nikolas Mittag, Nina Roussille, Amelie Schiprowski, Daphné Skandalis, Jason Sockin, Arne Uhlendorff, Wilbert van der Klauuw, Jacob Wegmann and participants at the annual meeting of the German Economic Association (VfS) 2022, the Berlin Network of Labor Market Research (BeNA) Winter Workshop 2022, the European Society for Population Economics (ESPE) 2023, the European Association of Labor Economics (EALE) 2023, and the Society for Labor Economists (SOLE) 2024 for their valuable comments and suggestions. We are grateful to the Institute of Labor Economics (IZA) and the Research Data Center (FDZ) of the Federal Employment Agency at the Institute of Employment Research (IAB) for access to the survey and administrative data from the IZA/IAB Evaluation Dataset. Marco Caliendo gratefully acknowledges funding from the German Research Foundation (Deutsche Forschungsgemeinschaft, DFG, project 405629508).}} 

	\author{
	\begin{tabular}{ccc}
		\textbf{Marco Caliendo}\thanks{University of Potsdam, Berlin School of Economics, IZA, DIW, IAB; \texttt{caliendo@uni-potsdam.de}} &&
		\textbf{Robert Mahlstedt}\thanks{University of Copenhagen, IZA; \texttt{robert.mahlstedt@econ.ku.dk}} \\
		\textbf{Aiko Schmeißer}\thanks{University of Potsdam, Berlin School of Economics; \texttt{schmeisser@uni-potsdam.de}} &&
		\textbf{Sophie Wagner}\thanks{University of Potsdam; \texttt{sopwagner@uni-potsdam.de}}\\
		\newline
		\newline
		\\[-3.0ex]
		\\
	\end{tabular}%
	\\}

\date{
August 2024
	\\ 
%	\textit{PRELIMINARY VERSION, PLEASE DO NOT CIRCULATE.} 
	\vspace{-5mm}
}

\maketitle

\begin{abstract}%\textbf{Initial Abstract/Idea}
	
	\noindent We study the accuracy of job seekers' wage expectations by comparing subjective beliefs to objective benchmarks using linked administrative and survey data. Our findings show that especially job seekers with low objective earnings potential and those predicted to face a penalty compared to their pre-unemployment wage display overly optimistic wage expectations. Moreover, 	 wage optimism is amplified by increased job search incentives and job seekers with overoptimistic wage expectations tend to overestimate their reemployment chances. We discuss the labor market implications of wage optimism, as well as the role of information frictions and motivated beliefs as sources of overoptimism.  \\

	\vspace{0.5cm}
	

\end{abstract}
% No page number on title page
\thispagestyle{empty}
\clearpage
\newpage
\setcounter{page}{1}

\section{Introduction}

The job search process is challenging, with unemployed workers facing substantial uncertainty regarding their job finding prospects \citep[see, e.g.,][]{Spinnewijn2015,Balleer2021,MuellerSpinnewijnTopa2021,ABGR2022} and their potential job matches \citep[see, e.g.,][]{KM2016,BKM2018,jager2022worker,sockin2023}. This, in turn, can distort their decision-making and may increase the risk of long-term unemployment. However, despite increasing evidence of systematic biases in job seekers' beliefs \citep[see][for an overview]{mu/sp/2022}, our understanding of the factors driving these misperceptions and the specific groups of job seekers most affected by them remains limited.

In our study, we examine the accuracy of job seekers' expectations about their earnings upon reemployment and analyze heterogeneity in the extent to which they exhibit overly optimistic or pessimistic beliefs about their potential earnings. We explore a unique combination of survey and administrative data on job seekers in Germany. The large-scale survey provides insights into the perceptions of more than 5,000 newly unemployed workers about their monthly wages upon reemployment. In addition, we use administrative records to establish objective benchmarks for their true earnings potential based on the realized wages of comparable workers with similar characteristics and labor market biographies. To approximate job seekers' objective wage potential, we account for these factors in flexible LASSO regressions.

While job seekers, on average, overestimate their reemployment wages by about 17\%, the comparison of subjective beliefs and objective benchmarks allows us to show that there is significant heterogeneity in the degree of overoptimism along two important dimensions. To begin with, we find that especially job seekers with lower objective earnings potential tend to be overly optimistic about their reemployment wages. Those positioned in the lowest decile of the objective benchmark distribution overestimate their potential wages by approximately 36\%, whereas the level of overoptimism is comparatively modest at around 6\% among individuals in the top decile of the distribution. 

In comparing job seekers' wage expectations to their pre-unemployment wages, we show that expected wage changes are more tightly compressed around zero compared to the distribution of objectively predicted wage changes. Hence, job seekers anchor their wage expectations more closely to their pre-unemployment wages than justified by realized wage growth of comparable workers. At the same time, this anchoring effect is asymmetric: individuals who can expect a wage increase compared to their previous salary have accurate beliefs, while those facing a potential wage decrease anchor their expectations too strongly to their pre-unemployment wage. This finding indicates that job seekers do not sufficiently account for the potential scarring effects of unemployment when forming their wage expectations.\footnote{Existing evidence suggests that the experience of unemployment is associated with wage penalties upon reemployment \citep[see, e.g,][]{arulampalam2001unemployment,gregory2001unemployment}. Consistent with this notion, our objective benchmarks indicate that job seekers' average wage potential decreases by about 12\% compared to their pre-unemployment wage.}

%Moreover, we observe considerable heterogeneity in the accuracy of job seekers' wage expectations concerning their personal characteristics. For example, we find that -- conditional on their objective earnings potential -- men and high-skilled workers tend to overestimate their wages more strongly than women and low-skilled workers.\footnote{The observed patterns align with earlier evidence suggesting gender differences in perceived labor market prospects \citep[see, e.g.,][]{CLM2017,reuben2017preferences,cortes2021gender} and indicating that higher levels of education are often linked to individuals being more overconfident \citep{bhandari2006demographics,trejos2019overconfidence}.}

Upon further investigating the nature of job seekers' optimistic beliefs, we document three additional results. First, there is considerable heterogeneity in the accuracy of job seekers' wage expectations concerning their personal characteristics. For example, we find that -- conditional on their objective earnings potential -- men and high-skilled workers tend to overestimate their wages more strongly than women and low-skilled workers.\footnote{The observed patterns align with earlier evidence suggesting gender differences in perceived labor market prospects \citep[see, e.g.,][]{CLM2017,reuben2017preferences,cortes2021gender} and indicating that higher levels of education are often linked to individuals being more overconfident \citep{bhandari2006demographics,trejos2019overconfidence}.} Second, we explore repeatedly elicited wage expectations, demonstrating that the level of overoptimism remains persistent among those who are still searching for a job until about one year after becoming unemployed. This pattern aligns with previous evidence \citep{KM2016,MuellerSpinnewijnTopa2021} and suggests that those facing challenges in securing a job are hesitant to adjust their wage expectations despite the feedback they receive during the job search process. Third, we provide evidence that job seekers believe to have control over their reemployment wage by submitting a greater number of applications. 

To establish the latter, we exploit quasi-exogenous variation in the incentives of unemployed workers to search for jobs. Specifically, we leverage regional differences along the administrative borders of local employment agency (LEA) districts, where job seekers face varying risks of being subject to punitive benefit sanctions if they provide an insufficient number of job applications. We demonstrate that the variation we exploit, in terms of local sanction risk, is independent of job seekers' characteristics or other aspects of LEAs' policy style, allowing us to estimate the causal effects of job seekers' perceived incentives for applying to and accepting jobs. We find that a stricter sanction regime encourages job seekers to intensify their search and simultaneously fosters greater optimism regarding their earnings potential. This indicates that job seekers believe they can control their reemployment wages, likely because they expect that sending out more applications will increase the likelihood of receiving higher-wage offers. Moreover, this control belief appears to outweigh any direct effect of the sanction risk causing job seekers to become less selective.

Holding overly optimistic wage expectations can have severe consequences for job seekers' labor market integration, as it may lead them to be excessively selective and reject offers more frequently than warranted, thereby prolonging unemployment \citep[see also][]{dubra2004optimism,conlon2018labor,MuellerSpinnewijnTopa2021,cortes2021gender}. A natural policy response would be to provide job seekers with information to correct their misperceptions. However, such a policy can have unintended consequences, especially given our finding that wage optimism is most prevalent among job seekers with low objective earnings potential and among those predicted to face a wage penalty compared to their previous salary. Correcting these misperceptions may discourage individuals, potentially reducing their effort or causing them to give up on finding a job altogether.\footnote{The existing experimental literature provides evidence supporting both the selectivity and motivation channels. For instance, \cite{jiang2023information} provide major-specific salary information to college graduates before entering the labor market and observe a negative relationship between reservation wages and subsequent job placement. Conversely, recent studies in South Africa and Uganda document that interventions reducing over-optimism depress job search and job finding \citep{bandiera2023search,banerjee2023learning}. More distantly related, various studies have investigated the effects of informing unemployed workers about potentially promising job matches \citep{BKM2018,AltmannEtAl2022,behaghel2022encouraging,dhia2022can,belot2022long} or the search process in general \citep{Altmann2018}, which can have positive effects on job seekers' labor market integration.} 

%\footnote{In line with this mechanism,  Moreover, \cite{jager2022worker} provide employed workers with information about their outside options, that is, the average wage of workers with similar characteristics in the same labor market, leading treated individuals to revise their wage expectations, as well as their job search and wage negotiation intentions. Finally, }

% These concerns are particularly serious because many job seekers in Germany have an objective earnings potential that is close to their monthly unemployment benefit level. For these workers, maintaining an optimistic outlook on their future labor market prospects may fulfill essential psychological needs by enhancing their motivation and helping them overcome self-control problems \citep{Be/Ti/2002,Be/Ti/2004,benabou2016mindful}.

With this in mind, the final part of our analysis provides descriptive evidence on the labor market implications of wage optimism. The matched survey-administrative data enable us to examine the extent to which job seekers' belief accuracy predicts their search behavior, their realized wages, as well as their perceived and actual job finding rates. The descriptive patterns align with the notion that optimism about wages motivates job seekers to actively search and be selective in accepting offers. Specifically, there is a positive relationship between wage optimism and the number of job applications, as well as realized wages upon finding employment. At the same time, we document a wedge between the perceived and actual job finding rates for increasing levels of wage optimism. On the one hand, job seekers who are most optimistic about their potential wages also report the highest perceived chances of finding a job. On the other hand, job seekers' actual prospects of finding a job decline as their wage optimism rises. This suggests that the more optimistic workers are about the wages they can earn upon reemployment, the more likely they are to overestimate their job finding prospects. This pattern aligns with the idea that optimistic job seekers who receive wage offers lower than their expectations tend to be excessively selective in accepting job offers. This, in turn, may prolong unemployment beyond their initial expectations.% \citep[see also][]{dubra2004optimism,conlon2018labor,MuellerSpinnewijnTopa2021}.\footnote{Consistent with our findings, \cite{} document gender differences in overoptimism and risk aversion among recent graduates, which contribute to women accepting job offers with lower pay than men.}

What causes the overall optimism of unemployed workers and the heterogeneity in their beliefs? It is often argued that job seekers have incomplete information about the job finding process and learn about their labor market prospects during job search \citep{burdett1988declining,conlon2018labor,gonzalez2010equilibrium}. Aligning with this notion, our findings suggest that job seekers with greater unemployment experience tend to hold more accurate earnings expectations. However, an alternative view is that misperceptions can arise from the way individuals process the information available to them. The literature on motivated reasoning suggests that individuals may hold optimistic beliefs as they derive direct utility from a positive self-image or to enhance their motivation \citep{Be/Ti/2002,Be/Ti/2004,benabou2016mindful}. Several of our findings speak to the empirical relevance of these ideas for the accuracy of job seekers' wage expectations. In particular, we find higher levels of overoptimism among individuals with the lowest objective earnings potential, especially among workers who are predicted to experience a wage decline compared to their past salary. This group of workers may have a heightened desire for motivated beliefs. Furthermore, job seekers who remain unemployed for an extended period are reluctant to revise their wage expectations downwards although they should have received signals about their earnings potential. This aligns with the idea that individuals may deliberately forget or suppress negative feedback to maintain their optimistic beliefs \citep{zimmermann2020dynamics}. Finally, job seekers who are encouraged to search more intensively due to extrinsic incentives hold more optimistic wage expectations, yet this does not translate into higher realized wages for them. These individuals may adopt a more optimistic outlook to motivate themselves to comply with their search requirements.

%Against this backdrop, one may expect that providing job seekers with information about their objective earnings potential may reduce their tendency to hold excessively optimistic wage expectations.\footnote{Related to this idea, various studies have investigated the effects of informing unemployed workers about potentially promising job matches \citep{BKM2018,AltmannEtAl2022,behaghel2022encouraging,dhia2022can,belot2022long} or the search process in general \citep{Altmann2018}, which can have positive effects on job seekers' labor market integration. Moreover, \cite{jager2022worker} provide employed workers with information about their outside options, that is, the average wage of workers with similar characteristics in the same labor market, leading treated individuals to revise their wage expectations, as well as their job search and wage negotiation intentions.}

%Our further results are also consistent with this idea of motivated beliefs. 

By establishing objective benchmarks for the subjective wage expectations of unemployed workers,  we contribute to a growing body of literature demonstrating job seekers' average tendency to be overly optimistic about their job finding prospects \citep{Spinnewijn2015,Balleer2021, MuellerSpinnewijnTopa2021,mu/sp/2022,Berg2022,abebe2023matching} and their reluctance to update their wage expectations over time \citep{KM2016,conlon2018labor,drahs2018job}.\footnote{Using a similar approach, \cite{jager2022worker} investigate the beliefs of employed workers regarding their outside options. Consistent with our findings, they discover that employed workers strongly anchor their wage expectations to their current salary. This anchoring effect is even more pronounced than what we observed among unemployed workers in our study.} In this context, our approach enables us to extend the literature in two dimensions. First, we can document significant heterogeneity in the accuracy of job seekers' wage expectations, with the highest levels of optimism observed among workers with poor labor market prospects. Second, it allows us to examine the empirical relationship between wage optimism, job search and labor market outcomes. We find that optimistic wage expectations are related to higher perceived but lower actual job finding rates, indicating that overly optimistic beliefs contribute to excessive selectivity in accepting job offers. %\citep[see, e.g.,][]{MuellerSpinnewijnTopa2021}. 

Additionally, our study sheds light on how job seekers' incentives to search for employment influence their behavior and beliefs. In doing so, we contribute to a limited body of research that examines the response of job search behavior to changes in the benefit environment. Aligning with existing studies investigating the stringency of sanction regimes \citep{ArniVanDenBergLalive2022} and the generosity of UI benefit payments \citep[see, e.g.,][]{marinescu2017general,lichter2021benefit}, we find that more restrictive policy regimes encourage unemployed workers to search for jobs more intensively. In addition, our finding that an enhanced sanction risk fosters greater wage optimism among unemployed workers challenges the notion that a stricter regime makes job seekers less selective in their job choices. Rather, the intensified search makes job seekers more optimistic about reemployment wages, likely because the increased search intensity raises the chances of receiving high-paying offers. This mechanism may contribute to the observation that, in various settings, reservation wages do not respond to variation in benefit payments or changes in the benefit rules as predicted by standard job search theory \citep{schneider2008effect,KM2016,le2019unemployment,lichter2021benefit}.\footnote{Using French administrative data on reservation wages and changes in UI rules, \cite{le2019unemployment} estimate a null effect of the potential benefit duration on reservation wages, which is consistent with the estimates by \cite{lichter2021benefit} based on German survey data. Similarly, \cite{KM2016} cannot reject that the elasticity of reservation wages to benefit levels in the U.S. is equal to zero, while \cite{schneider2008effect} finds no significant effect of imposed benefit sanctions on the reservation wages of unemployed workers in Germany.} At the same time, there is no indication that job seekers' increased wage optimism, which comes with the higher search intensity, is warranted since realized wages upon reemployment tend to be (insignificantly) lower when job seekers are subject to a more restrictive sanction regime. %This aligns with previous evidence suggesting that benefit sanctions often lead job seekers to eventually accept lower-quality jobs \citep{arni2013effective,van2014monitoring,van2019evaluating}. Hence, it is conceivable that job seekers facing an elevated risk of sanctions only lower their wage expectations as time progresses.

%restrictive policy regimes, that is, caseworkers who apply greater pressure on job seekers and tend to issue benefit sanctions more frequently. \textbf{XXX add brief overview of literature and connection to our findings XXX}

The remainder of this paper proceeds as follows. In the next section, we discuss our empirical setting, while Section \ref{sec:results} presents empirical evidence on the accuracy of job seekers' wage expectations. Section~\ref{sec:discussion_implications} discusses the labor market implications of wage optimism in a job search framework and provides descriptive empirical evidence. Finally,  Section~\ref{sec:concl} concludes.

\section{Empirical Setting} \label{sec:data}

To examine the accuracy of job seekers' wage expectations, we build on different complementary data sources providing information on unemployed workers in Germany. To begin with, we rely on a large-scale survey involving 17,400 workers who became unemployed between June 2007 and May 2008 and were eligible for unemployment insurance (UI) benefits \citep[see][]{ACKZ14}. The first interview was conducted within 7 to 14 weeks after entering unemployment, followed by a second interview wave 12 months later. The survey encompasses detailed data on socio-demographic characteristics, personality traits, job search behavior, and, notably for our study, subjective beliefs about labor market prospects, especially wages upon reemployment.

In addition to the survey, we leverage administrative records to access highly reliable data regarding job seekers' actual labor market outcomes and their employment histories prior to unemployment. We use the administrative data for two purposes. First, we can directly link the survey information with the administrative records at the individual level for about 87\% of the survey respondents \citep{FDZ2017}. Second, we incorporate administrative information from a larger sample of unemployed workers to establish objective benchmarks for job seekers' earnings potential. Importantly, both datasets consist of individuals randomly sampled from the same population of unemployed workers. This ensures that job seekers within the different datasets can be directly compared and that we have access to similar information regarding their labor market biographies.\footnote{The \textit{IZA/IAB Administrative Evaluation Dataset} offers administrative data for a 4.7\% random sample of individuals who entered unemployment between 2001 and 2008 \citep{CFA11,FDZ2015}, while the \textit{IZA Evaluation Dataset Survey} provides survey data for a representative subset of individuals who entered unemployment between June 2007 and May 2008. Furthermore, the \textit{IZA/IAB Linked Evaluation Dataset} combines survey and administrative data, linking them for 87\% of the survey respondents.}

\subsection{Subjective wage expectations and objective benchmarks}\label{sec:data_sub}

The survey elicits job seekers' beliefs about the monthly net salary they expect to receive upon starting a new job, using the following question:
\begin{quote}
	 ``Now, I am interested in the salary you anticipate receiving in your next job. What is your expected monthly net income in \euro?''
\end{quote} 
The question is asked during the initial survey interview, which takes place 7 to 14 weeks after entry into unemployment, and is directed at all individuals who are still unemployed at this stage and are actively searching for a job. Moreover, our analysis focuses exclusively on individuals who previously held full-time positions to minimize the influence of variation in working hours on monthly wage expectations. This results in an estimation sample of 5,376 survey respondents who can be linked to the administrative records. 

A straightforward approach to analyze the accuracy of individuals' wage expectations is to compare subjective beliefs to actual wage outcomes of the same individual. However, realized wages are only observed for those workers who actually start a new job, and they are influenced by their prior beliefs. This complicates the analysis of how belief inaccuracies impact long-term unemployment. Moreover, realized wages represent only one draw from the wage offer distribution and are affected by luck and unforeseeable labor demand shocks, which individuals cannot account for when reporting their subjective beliefs. As a consequence, an individual's realized wage does not necessarily reflect their average earnings potential at the time their beliefs were elicited. To address these concerns we adopt an approach where we estimate objective benchmarks for job seekers' earnings potential based on the realized wages of comparable individuals in similar situations. At the same time, we acknowledge that realized wages may offer insight into the relevance of unobserved heterogeneity at the individual level. Therefore, we compare the objective benchmarks with job seekers' realized wages, finding that prediction errors do not systematically vary with the objective benchmark.

\begin{comment}
\AS{
	\bi
	\item Develop arguments in more detail + (add to intro?)
	\item Emphasize more: In a random job search model wages are random draws from a wage offer distribution (conditional on being larger than the reservation wage) such that for a given individual realized wages are also determined by luck and do not need to reflect the mean of the wage distribution
	\item Add: Realized wages are only observed among the reemployed
	\ei
}
\end{comment}

%Our observation that job seekers expect labor market outcomes above their actual realizations aligns with previous evidence from \cite{Spinnewijn2015}, \cite{balleer2021effects}, and \cite{MuellerSpinnewijnTopa2021} indicating that a majority of job seekers tend to hold overly optimistic beliefs about their employment prospects. However, drawing conclusions about the accuracy of subjective expectations at a more fine-grained level it is notoriously challenging for two reasons. First, at the individual level, realized outcomes represent only one instance of the objective ex-ante distribution of wages and realizations might be affected by unforeseeable labor demand shocks that individuals cannot be aware of when reporting their subjective beliefs. 

We use administrative data from a larger sample of 84,617 workers who became unemployed between January 2005 and May 2007 (see Appendix~\ref{app:data} for additional information regarding the prediction of objective benchmarks). This time period was chosen to avoid any overlap with the survey sample, ensuring that the objective predictions are not influenced by the beliefs and behaviors of survey respondents. As a robustness check, we also employ a random sample comprising 80\% of all entries into unemployment between June 2007 and May 2008, which aligns with the survey period. Detailed summary statistics for both samples can be found in Appendix Table~\ref{tab:sumstat_led_aed}. Importantly, average reemployment wages are very similar across cohorts, indicating that macroeconomic trends, such as the financial crisis, do not undermine the validity of our objective benchmarks. In order to ensure comparability with the survey sample, we apply similar restrictions to the administrative data. Specifically, we focus on newly unemployed individuals who are eligible for unemployment insurance (UI) benefits and were previously employed in non-subsidized full-time positions for at least three months. Moreover, we restrict the sample to job seekers who have not found regular employment within three months after entry, which is the average time until the first interview of the survey.  

We employ flexible LASSO regressions to predict reemployment wages, accounting for a comprehensive set of pre-determined covariates available in the matched survey-admin data. This includes socio-demographic characteristics, information on the last job before unemployment, labor market history over the past ten years, and local labor market characteristics. The dependent variable is the first monthly salary received in a regular job within 24 months after entry into unemployment, and we test the robustness of our findings using different time horizons.\footnote{Specifically, we use wages of individuals reemployed within nine months of unemployment, which is, on average, six months after the initial interview. This is motivated by the fact that job seekers answered the question about their wage expectations shortly after discussing their anticipated likelihood of finding a job during the next six months, suggesting a consistent time frame for wage expectations.}
To compare the objective benchmarks with subjective wage expectations, we convert the realized wages recorded in administrative records from gross to net terms by deducting social security contributions and income taxes (see Appendix~\ref{app:data} for details). 

To evaluate the quality of the benchmarks, we estimate the out-of-sample $R^2$ by regressing realized wages on predicted wages using distinct test datasets, i.e., samples not utilized during the prediction generation process. As shown in Appendix Table~\ref{tab:performance}, we find values of $R^2$ within the range of 0.48 to 0.53, suggesting that we are equipped with meaningful objective benchmarks for individuals' wages.\footnote{These figures align with other studies that predict wages of German workers and report out-of-sample $R^2$ values ranging from 0.4 to 0.5 when accounting for worker characteristics \citep{Card2013,jager2022worker}.} Further supporting this notion, Appendix Table~\ref{tab:performance_objVSsub} reveals that the objective benchmarks derived from wages of comparable workers exhibit greater predictive power for survey respondents' realized wages compared to their own subjective wage expectations. %\deleted{Specifically, we find a coefficient of 0.95 when regressing realized wages and objective predictions in our survey sample.}
%For a comparison to other wage predictions in Germany, see \cite{jager2022worker} who obtain an out-of-sample $R^2$ of 0.40 for wage changes after firm-to-firm transitions and \cite{Card2013} who find that basic Mincer controls (education, experience, industry, occupation) yield an $R^2$ of 0.50 for wages in the overall working population.} 
%Jason: I would have a reference here to some benchmark R2 for predicting wages.  For instance, what would AKM get you for an R2 on german admin data. I think Card et al. have done this, I think its like 90\%, but they may also have the R2 without firm FE, something like that.

%\footnote{\AS{In Appendix Table \ref{tab:performance_objVSsub}, . This already suggests that measurement error in the objective predictions is unlikely to fully explain the observed slope between expected wages and objective predictions.}}

\subsection{Summary statistics}

\begin{figure}[h!]
	\caption{Distribution of subjective wage beliefs and objective benchmarks}  \label{fig:dist}
	\centerline{
		\begin{threeparttable}
			\begin{footnotesize}
				\begin{tabular}{cc}
					\textbf{A. Subjective wage beliefs} & \textbf{B. Objective benchmarks}\\
					\includegraphics[width=0.45\linewidth]{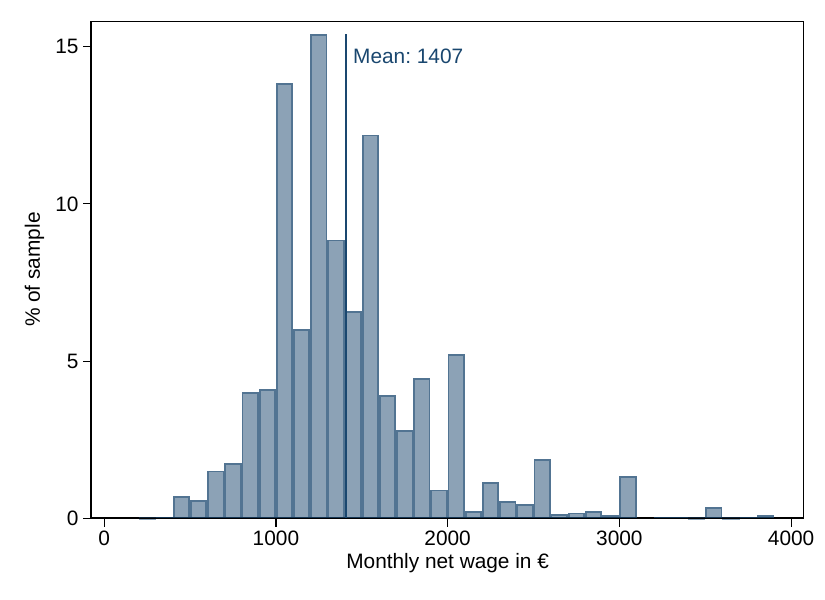} &\includegraphics[width=0.45\linewidth]{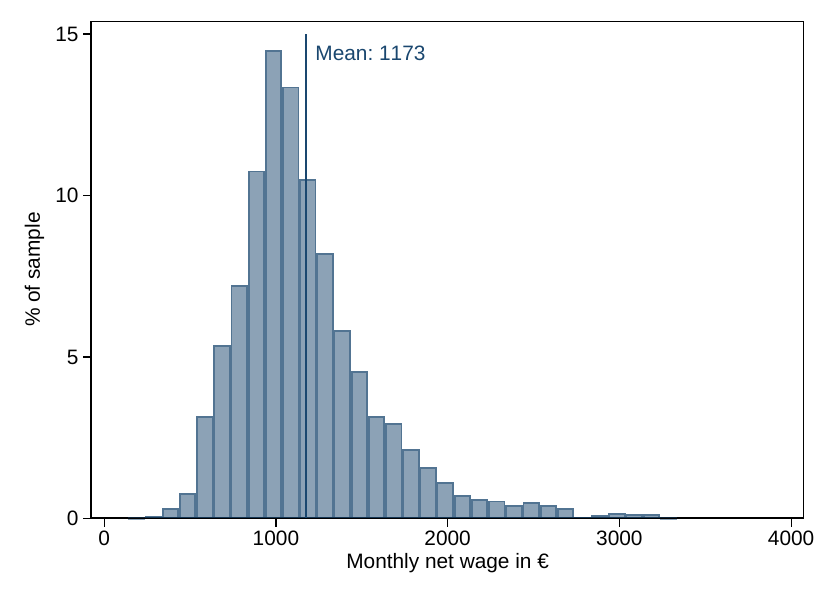} \\ 
				\end{tabular}
			\end{footnotesize}
			\begin{tablenotes} \scriptsize
				\item \textit{Note:}
				The figure shows the distribution of subjective beliefs (Panel~A) and objective benchmarks (Panel~B) for individuals' monthly net income upon reemployment among the sample of survey respondents ($N = 5,376$).
				Objective benchmarks are generated from realized outcomes of similar individuals observed in the administrative records (see Section \ref{sec:data_sub} for details). Panel~A excludes individuals with an expected reemployment wage greater than 4,000\euro~($<1\%$ of sample).
			\end{tablenotes}
	\end{threeparttable}}
\end{figure}

Figure~\ref{fig:dist} shows the distributions of subjective wage expectations and objective benchmarks. The average expected net income is 1,407\euro~per month, which is substantially greater than the average income that workers could reasonably expect. For the average job seeker in our sample, the objective benchmarks suggest a monthly net wage of only 1,173\euro~. This closely aligns with the average realized wage of 1,190\euro~per month that we observe among the survey respondents (see Panel~A of Appendix Table~\ref{tab:sum_stat}). 

In addition to job seekers' wage expectations, we also explore information on individuals' perceived and actual job finding prospects. The perceived chances of reemployment are elicited over a six-month horizon and responses are provided using four options: ``very likely'', ``likely'',``unlikely'', or ``very unlikely''.\footnote{Specifically, survey respondents answer the following question: ``When you think of the future, how likely is it from your perspective that you will find a job within the next six months?''} Overall, unemployed workers in our survey sample tend to be remarkably optimistic about their reemployment prospects. In particular, 89\% of the survey population consider themselves ``likely'' or ``very likely'' to find a job within six months, while only 56\% actually do so (see Panel~B of Appendix Table~\ref{tab:sum_stat}). This observation is in line with existing evidence from, e.g., \cite{Spinnewijn2015}, \cite{Balleer2021}, and \cite{MuellerSpinnewijnTopa2021}, suggesting that a majority of job seekers hold overly optimistic beliefs about their job finding probabilities. 

In our main analysis, our focus is on the accuracy of job seekers' wage expectations because respondents' subjective beliefs regarding their job finding prospects are captured on an ordinal scale with only four values. This complicates direct comparison with objective benchmarks for perceived job finding probabilities.
However, in Section~\ref{sec:implications_results}, we further investigate the empirical relationship between wage optimism and both perceived and actual job finding rates.

 %However, in Appendix Figure~\ref{fig:accuracy_jf}, we present a preliminary analysis indicating a similar pattern for reemployment and wage expectations. \textbf{RM: Add main figure (i.e. Figure~\ref{fig:subjective_objective}) for reemployment expectations in Appendix.} 

\section{Empirical Evidence on the Accuracy of Wage Expectations} \label{sec:results}

In this section, we compare job seekers' subjective wage expectations to the objective benchmarks. This enables us to uncover heterogeneity in the accuracy of job seekers' wage expectations, taking into account their objective earnings potential (see Section~\ref{sec:subjective_objective}) and individual background characteristics (see Section~\ref{sec:determinants}). Moreover, we investigate the extent to which job seekers anchor their wage expectations to their previous salary (see Section~\ref{sec:anchor}) and examine how they revise their subjective expectations over the course of their unemployment spell (see Section~\ref{sec:update}). Finally, we explore whether job seekers believe to have control over their reemployment wages by altering their search behavior (see Section~\ref{sec:incentives}).

\subsection{Heterogeneity by objective earnings potential}\label{sec:subjective_objective}

To begin with, we consider the relationship between subjective expectations and objective benchmarks based on the following regression model \citep[see][]{jager2022worker}:
\begin{equation}
	S_i = \beta_0 + \beta_1 \widehat{O_i} + \epsilon_i \label{eq:reg_bias}
\end{equation}	
where $S_i$ denotes the subjective belief of job seeker $i$ about their reemployment wage and $\widehat{O}_i$ refers to the corresponding objective prediction. The intercept $\beta_0$ captures inaccuracies in beliefs that are common to all job seekers, while the slope parameter $\beta_1$ describes how strongly beliefs respond to variation in objective benchmarks. In this context, we can think about different scenarios depending on the values of $\beta_0$ and $\beta_1$.

First, when $\beta_1=1$ and $\beta_0=0$, individuals' expectations perfectly correspond to the objective benchmark, which indicates \textit{accurate beliefs} throughout the distribution. Second, when $\beta_1=1$ and $\beta_0\neq0$, job seekers' beliefs are subject to \textit{homogeneous inaccuracies}, that is, individuals are overly optimistic ($\beta_0>0$) or pessimistic ($\beta_0<0$), but job seekers with different objective predictions share the same degree of optimism or pessimism. Lastly, when $\beta_1\neq1$, beliefs do not exhibit a one-to-one response to objective variation, resulting in \textit{heterogeneous inaccuracies}. In particular, $\beta_1<1$ means that beliefs do not adjust sufficiently strongly to changes in objective predictions, which implies that overoptimism is more pronounced among job seekers with a relatively low objective wage potential. 

\begin{figure}[h]
	\caption{Comparison of subjective beliefs and objective benchmarks \label{fig:subjective_objective}}
	\centerline{
		\begin{threeparttable}
	\begin{footnotesize}
		\begin{tabular}{cc}
			\textbf{A. Relationship}  & \textbf{B. Distribution} \\
%					\textbf{subjective beliefs and objective benchmarks}\\
					\includegraphics[width=0.45\linewidth]{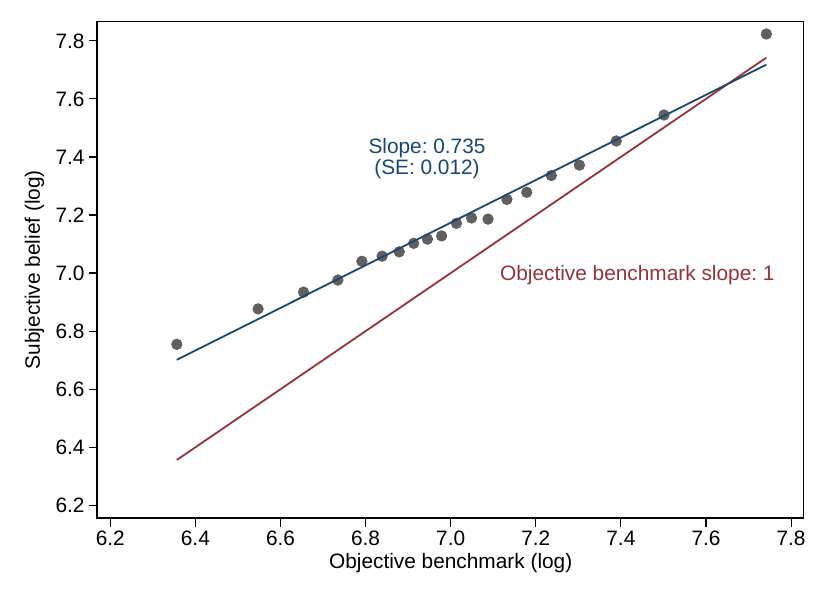} &
					\includegraphics[width=0.45\linewidth]{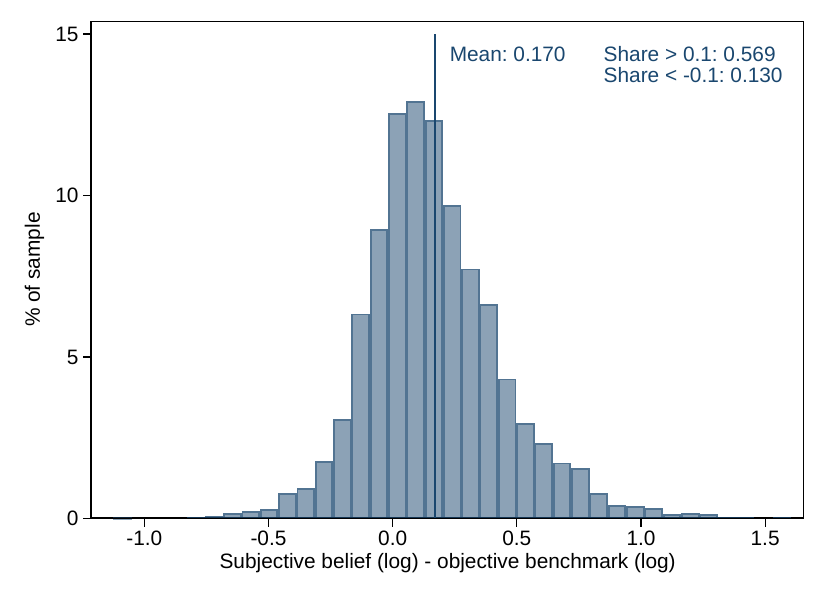}\\
				\end{tabular}
			\end{footnotesize}
		\begin{tablenotes} \scriptsize
\item \textit{Note:} The figure compares subjective beliefs and objective benchmarks for reemployment wages among the sample of survey respondents ($N=5,376$). Panel~A illustrates Equation~\eqref{eq:reg_bias} by plotting 
individuals' subjective beliefs against objective benchmarks (each of the 20 dots represents a ventile of the distribution) and the red line represents the accurate benchmark if subjective expectations were to equal objective predictions. The estimated slope is significantly different from 1 (p-value $<$ 0.001). Panel~B illustrates the distribution of individual-level log differences between subjective and objective benchmarks. 
			\end{tablenotes}
	\end{threeparttable}}
\end{figure}

Panel A of Figure~\ref{fig:subjective_objective} displays a binned scatter plot based on Equation \eqref{eq:reg_bias}. It illustrates the prevalence of overoptimism regarding reemployment wages across all levels of the corresponding objective distribution. In each of the 20 bins, the average expected wage exceeds the corresponding objective benchmark. In other words, both job seekers with low and high objective earnings potential tend to overestimate their reemployment wage compared to our objective benchmarks. At the same time, the magnitude of job seekers' optimism varies across the objective distribution. Our analysis reveals a slope coefficient of $\widehat{\beta}_1=0.74$ (SE: 0.01) compared to an objective benchmark slope of one. This suggests that beliefs do not adequately respond to variations in the objective wage potential. As a result, job seekers with lower predicted wages exhibit a greater tendency to overestimate their earnings potential compared to those with higher objective predictions. To be precise, individuals within the bottom decile of the objective benchmark distribution overestimate their earnings potential by approximately 36\%, whereas the overoptimism is only about 6\% among individuals in the top decile of the distribution.\footnote{In absolute terms, this corresponds to an overestimation of wages by 323\euro~in the bottom decile and by 214\euro~in the top decile of the objective wage distribution.}

\begin{figure}[h!]
	\caption{Relation between subjective belief uncertainty and objective benchmarks \label{fig:subjectiveSD_objective}}
	\centerline{
		\begin{threeparttable}
			\begin{footnotesize}
				\begin{tabular}{cc}
					\includegraphics[width=0.7\linewidth]{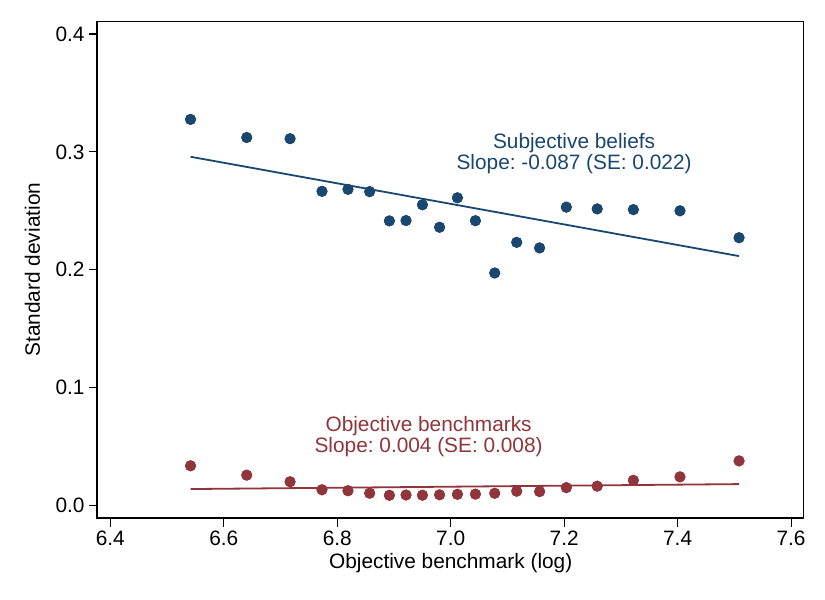}
				\end{tabular} 
			\end{footnotesize}
			\begin{tablenotes} \scriptsize
				\item \textit{Note:}
				The figure plots the standard deviation of individuals' subjective wage beliefs (blue line) and of objective benchmarks (red line) against the level of objective benchmarks. Each of the 20 dots shows the standard deviation that is calculated within a ventile of the objective benchmark distribution. The slope coefficient is significantly different from 0 for subjective beliefs (p-value $= 0.001$) and not significant for objective benchmarks (p-value $=0.591$). The sample is trimmed at the top and bottom 5\% of objective benchmarks. $N = 4,840$.
			\end{tablenotes}
	\end{threeparttable}}
\end{figure}

\paragraph{Disagreement among job seekers:} Besides the heterogeneity in the difference between subjective beliefs and objective benchmarks, we also consider the extent to which job seekers with similar objective benchmarks disagree in their wage expectations. To that end, we measure the standard deviation of job seekers' beliefs within each bin of the objective benchmark distribution. Figure~\ref{fig:subjectiveSD_objective} shows that the dispersion in subjective beliefs tend to be larger among individuals at the lower part of the objective benchmark distribution. This indicates that job seekers with a low earnings potential are not only the most optimistic but also the most uncertain about their reemployment wages.\footnote{We calculate the standard deviation of beliefs within equal-sized bins of the sample (see blue line in Figure~\ref{fig:subjectiveSD_objective}). While each bin has the same size, it may span different ranges of the objective benchmark distribution. Therefore, we also consider the standard deviation of objective benchmarks within each bin (see red line in Figure~\ref{fig:subjectiveSD_objective}), suggesting that the pattern is not driven by dispersion in objective benchmarks.} 

%Panel~A of Figure~\ref{fig:subjective_objective} shows a binned scatter plot of Equation \eqref{eq:reg_bias}. It becomes evident that optimisms about reemployment wages is widespread at all levels of the corresponding objective distribution, that is, the average expected wage in all 20 bins is higher than the corresponding objective benchmark.  At the same time, the magnitude of job seekers' optimism differs across the objective distribution. Compared to an objective benchmark slope of one, we estimate a slope coefficient of $\widehat{\beta}_1=0.78$ (SE: 0.01), suggesting that beliefs do not respond sufficiently to variation in the objective wage potential. This means that job seekers for whom we predict relatively low wages tend to overestimate their earnings potential to a greater extent than those with a high objective prediction. 

%\vspace{1em}

\paragraph{Robustness:} A series of robustness checks confirms our result that job seekers with lower objective earnings potential exhibit the highest levels of optimism. First, we consider the possibility that our results are driven by prediction errors in the objective benchmarks rather than by subjective beliefs. In Appendix Figure \ref{fig:obj_vs_realized}, we show that the objective benchmarks predict realized wages with an estimated slope coefficient very close to one, suggesting that predictions errors do not vary systematically over the objective benchmark distribution. Second, we restrict our sample to job seekers who find reemployment within 24 months, since objective benchmarks can only be estimated for those who ultimately become reemployed. Third, we vary the sample used to estimate the objective benchmark based on the administrative records. In particular, we rely on alternative training data including individuals who became unemployed in 2007 and 2008 (i.e. the same time period during which the survey was conducted). Fourth, we only include job seekers who find a job within nine rather than 24 months when calculating objective benchmarks. Fifth, we address concerns that the estimated slope coefficients may suffer from attenuation bias due to measurement error in the objective predictions. Therefore, we conduct an instrumental variable (IV) regression, where we use objective predictions from the alternative training data as an instrument for objective predictions from our baseline training data. This approach is similar to a split-sample IV measurement error correction \citep[see, e.g.,][]{DJPS2020, jager2022worker}. Sixth, we assess the role of rounding in subjective beliefs by also rounding the objective benchmarks to the nearest 50, 100, or 250\euro~values. Lastly, we impose two additional restrictions on the survey sample, considering only (i) individuals who search for and expect to find a full-time job and (ii) unmarried individuals. This enables us to examine whether differences in expected working hours (full-time) and measurement error arising from the conversion of gross to net wages (unmarried individuals) affect our estimates.\footnote{Notice that calculating social security contributions and income taxes is straightforward for unmarried individuals. Therefore, any measurement error arising from the conversion of gross to net wages is unlikely to impact the objective benchmarks for this group of individuals.} The results from all robustness checks are shown in Appendix Figure~\ref{fig:robustness} and Table~\ref{tab:subjective_objective_robust}. Across the various specifications, we obtain similar slope coefficients between 0.64 and 0.75.

%\textbf{AS: Should we show and reference Appendix Figure \ref{fig:obj_vs_subj_24} or is it sufficient to add the slope coefficient to the robustness table? RM: Maybe we can show Figure for different robustness specifications (perhaps the four or six most important ones). RM: To be discussed where to put this ``robustness'' paragraph.}

%\AS{Thus far, we have examined how the level of job seekers' subjective belief is related to the objective benchmark prediction, noting that job seekers with the lowest objective benchmarks, on average, overestimate their wages the most. In Appendix Figure~\ref{fig:subjectiveSD_objective}, we also consider the uncertainty around these averages by studying the variation in beliefs across job seekers. Specifically, we measure the standard deviation of job seekers' beliefs within each bin of the objective benchmark distribution. The results show that the dispersion in beliefs is somewhat larger within the group of individuals with the lowest predicted wages.} \AS{Thus, low-wage job seekers do not only appear to be the most optimistic but also the most uncertain about their earnings potential.}

%\AS{
%	\bi
%	\item Not sure whether Figure~\ref{fig:subjectiveSD_objective} looks very convincing. Standard deviations of objective benchmarks are also larger in the first and last bins. Better take it out? Or any other idea how to better analyze it? 
%	\ei
%}

\subsection{Determinants of optimism and pessimism} \label{sec:determinants}

Next, we consider deviations between subjective beliefs and objective benchmarks: $S_i - O_i$. As shown in Panel~B of Figure~\ref{fig:subjective_objective}, job seekers overestimate their wage potential, on average, by about 17\%. While a majority of about 57\% of the survey respondents expect their wage to be at least 10\% higher than the objective prediction, only 13\% underestimate their wage potential by more than 10\%.\footnote{In Appendix Table \ref{tab:subjective_objective_robust}, we report the mean deviation between subjective beliefs and objective benchmark, as well as the share of job seekers who over- and underestimate their wage potential for the alternative specifications explained above. Across all specifications, we find that overly optimistic wage expectations are widespread among unemployed workers.} Building upon this individual-level measure, we explore correlations between the accuracy of beliefs and job seekers' characteristics. To that end, we first regress subjective wage expectations on a set of covariates without accounting for the objective predictions (column~1 of Table~\ref{tab:correlates}). Afterwards, we analyze to what extent these differences in expectations are justified by heterogeneity in job seekers' actual wage potential. Therefore, we regress the deviation between subjective beliefs and objective benchmarks, $S_i - O_i$, on the covariates and control for objective benchmarks (column~2). Lastly, in columns (3) and (4), we distinguish between individuals who overestimate and underestimate their wage potential.\footnote{In particular, we consider the deviation between $S_i$ and $O_i$ and set negative (positive) values to zero and thus only exploit variation in positive (negative) deviations.}

%\footnote{\deleted{One should note that the covariates included in the regression model represent a subset of all individual-level characteristics that are explored to generate the objective benchmarks. This means the estimated correlations cannot be explained by heterogeneity in rational expectations.} \AS{We add objective benchmarks to the model to account for the result, presented in panel A of Figure~\ref{fig:subjective_objective}, that job seekers with a lower objective earnings potential are generally more overoptimistic about their earnings.}} 

\begin{table}[p]
	\centerline{
		\begin{threeparttable}
			\vspace{-55pt}
			\caption{Determinants of optimism and pessimism}\label{tab:correlates}
			\tabcolsep=0.25cm
			\begin{footnotesize}
				\begin{tabular}{lcccccccc}
					\hline\hline \\[-2.0ex]
					\\[-2.0ex]
					%& \mc{3}{c}{Job finding beliefs} && \mc{3}{c}{Wage beliefs}\\
					%\\[-2.0ex]
					%\cline{2-4}\cline{6-8}
					%\\[-2.0ex]		
					& $S_i$ & $S_i - O_i$ & $S_i - O_i$ & $S_i - O_i$   & $S_i - O_i$ & $S_i - O_i$ \\
					&  		&	  & Pos. values & Neg. values   &  Pos. values & Neg. values \\
					& (1) & (2) & (3) & (4) & (5) & (6)   \\
					\midrule %\\[-2.0ex]
					\mc{7}{l}{\textbf{Socio-demographic characteristics}}\\
Female      &      -0.172\sym{***}&      -0.086\sym{***}&      -0.067\sym{***}&      -0.019\sym{***}&      -0.064\sym{***}&      -0.019\sym{***}\\

            &     \raisebox{.7ex}[0pt]{\scriptsize(0.008)}         &     \raisebox{.7ex}[0pt]{\scriptsize(0.007)}         &     \raisebox{.7ex}[0pt]{\scriptsize(0.006)}         &     \raisebox{.7ex}[0pt]{\scriptsize(0.003)}            &     \raisebox{.7ex}[0pt]{\scriptsize(0.006)}         &     \raisebox{.7ex}[0pt]{\scriptsize(0.003)}         \\

	Age (ref. 16-24 years)\\
		\quad 25 - 34 years&       0.039\sym{***}&       0.022\sym{**} &       0.012         &       0.010\sym{**} &       0.014\sym{*}  &       0.010\sym{**}   \\
            &     \raisebox{.7ex}[0pt]{\scriptsize(0.011)}         &     \raisebox{.7ex}[0pt]{\scriptsize(0.010)}         &     \raisebox{.7ex}[0pt]{\scriptsize(0.008)}         &     \raisebox{.7ex}[0pt]{\scriptsize(0.004)}               &     \raisebox{.7ex}[0pt]{\scriptsize(0.008)}         &     \raisebox{.7ex}[0pt]{\scriptsize(0.004)}         \\
            
	\quad 35 - 44 years &       0.057\sym{***}&       0.014         &       0.005         &       0.009\sym{**} &       0.010         &       0.009\sym{**}  \\
            &     \raisebox{.7ex}[0pt]{\scriptsize(0.011)}         &     \raisebox{.7ex}[0pt]{\scriptsize(0.010)}         &     \raisebox{.7ex}[0pt]{\scriptsize(0.008)}         &     \raisebox{.7ex}[0pt]{\scriptsize(0.004)}               &     \raisebox{.7ex}[0pt]{\scriptsize(0.009)}         &     \raisebox{.7ex}[0pt]{\scriptsize(0.004)}         \\

\quad 45 - 55 years&       0.072\sym{***}&       0.027\sym{**} &       0.016\sym{*}  &       0.012\sym{***}&       0.026\sym{***}&       0.013\sym{***}\\
            &     \raisebox{.7ex}[0pt]{\scriptsize(0.012)}         &     \raisebox{.7ex}[0pt]{\scriptsize(0.011)}         &     \raisebox{.7ex}[0pt]{\scriptsize(0.009)}         &     \raisebox{.7ex}[0pt]{\scriptsize(0.004)}               &     \raisebox{.7ex}[0pt]{\scriptsize(0.009)}         &    \raisebox{.7ex}[0pt]{\scriptsize(0.004)}         \\

German citizen      &      -0.026\sym{*}  &      -0.046\sym{***}&      -0.032\sym{***}&      -0.014\sym{***}&      -0.037\sym{***}&      -0.014\sym{***}  \\
            &     \raisebox{.7ex}[0pt]{\scriptsize(0.015)}         &     \raisebox{.7ex}[0pt]{\scriptsize(0.014)}         &     \raisebox{.7ex}[0pt]{\scriptsize(0.012)}         &     \raisebox{.7ex}[0pt]{\scriptsize(0.005)}               &     \raisebox{.7ex}[0pt]{\scriptsize(0.012)}         &     \raisebox{.7ex}[0pt]{\scriptsize(0.005)}         \\

	\mc{4}{l}{Educational level (ref. no higher education)}\\
\quad Vocational certificate&       0.055\sym{***}&       0.018         &       0.003         &       0.015\sym{**} &      -0.003         &       0.014\sym{**} \\
            &     \raisebox{.7ex}[0pt]{\scriptsize(0.013)}         &     \raisebox{.7ex}[0pt]{\scriptsize(0.012)}         &     \raisebox{.7ex}[0pt]{\scriptsize(0.009)}         &     \raisebox{.7ex}[0pt]{\scriptsize(0.006)}                 &     \raisebox{.7ex}[0pt]{\scriptsize(0.010)}         &     \raisebox{.7ex}[0pt]{\scriptsize(0.006)}        \\

	\quad University degree&       0.266\sym{***}&       0.135\sym{***}&       0.118\sym{***}&       0.017\sym{***}&       0.107\sym{***}&       0.016\sym{**}\\
            &     \raisebox{.7ex}[0pt]{\scriptsize(0.016)}         &     \raisebox{.7ex}[0pt]{\scriptsize(0.016)}         &     \raisebox{.7ex}[0pt]{\scriptsize(0.013)}         &     \raisebox{.7ex}[0pt]{\scriptsize(0.006)}               &     \raisebox{.7ex}[0pt]{\scriptsize(0.013)}         &     \raisebox{.7ex}[0pt]{\scriptsize(0.007)}         \\

Married     &      -0.022\sym{***}&       0.002         &       0.015\sym{**} &      -0.013\sym{***}&       0.018\sym{***}&      -0.012\sym{***}\\
            &     \raisebox{.7ex}[0pt]{\scriptsize(0.008)}         &     \raisebox{.7ex}[0pt]{\scriptsize(0.008)}         &     \raisebox{.7ex}[0pt]{\scriptsize(0.007)}         &     \raisebox{.7ex}[0pt]{\scriptsize(0.003)}                &     \raisebox{.7ex}[0pt]{\scriptsize(0.007)}         &     \raisebox{.7ex}[0pt]{\scriptsize(0.003)}         \\

Any children   &      -0.026\sym{***}&      -0.012         &      -0.004         &      -0.007\sym{**} &      -0.003         &      -0.007\sym{**} \\
            &     \raisebox{.7ex}[0pt]{\scriptsize(0.009)}         &     \raisebox{.7ex}[0pt]{\scriptsize(0.008)}         &     \raisebox{.7ex}[0pt]{\scriptsize(0.007)}         &     \raisebox{.7ex}[0pt]{\scriptsize(0.003)}            &     \raisebox{.7ex}[0pt]{\scriptsize(0.007)}         &     \raisebox{.7ex}[0pt]{\scriptsize(0.003)}         \\
            
	East Germany   &      -0.081\sym{***}&      -0.015\sym{*}  &      -0.012\sym{*}  &      -0.003         &      -0.011         &      -0.002      \\
            &     \raisebox{.7ex}[0pt]{\scriptsize(0.008)}         &     \raisebox{.7ex}[0pt]{\scriptsize(0.008)}         &     \raisebox{.7ex}[0pt]{\scriptsize(0.007)}         &     \raisebox{.7ex}[0pt]{\scriptsize(0.003)}          &     \raisebox{.7ex}[0pt]{\scriptsize(0.007)}         &     \raisebox{.7ex}[0pt]{\scriptsize(0.003)}         \\
            
\mc{7}{l}{\textbf{Labor market history}}\\
Last wage (ln)&       0.325\sym{***}&       0.212\sym{***}&       0.176\sym{***}&       0.036\sym{***}&       0.169\sym{***}&       0.035\sym{***}\\

            &     \raisebox{.7ex}[0pt]{\scriptsize(0.013)}         &     \raisebox{.7ex}[0pt]{\scriptsize(0.013)}         &     \raisebox{.7ex}[0pt]{\scriptsize(0.011)}         &     \raisebox{.7ex}[0pt]{\scriptsize(0.004)}             &     \raisebox{.7ex}[0pt]{\scriptsize(0.011)}         &     \raisebox{.7ex}[0pt]{\scriptsize(0.004)}         \\

	Last job was quit   &      -0.025         &      -0.003         &       0.004         &      -0.007         &       0.001         &      -0.007         \\

            &     \raisebox{.7ex}[0pt]{\scriptsize(0.018)}         &     \raisebox{.7ex}[0pt]{\scriptsize(0.016)}         &     \raisebox{.7ex}[0pt]{\scriptsize(0.013)}         &     \raisebox{.7ex}[0pt]{\scriptsize(0.006)}                &     \raisebox{.7ex}[0pt]{\scriptsize(0.013)}         &     \raisebox{.7ex}[0pt]{\scriptsize(0.006)}         \\
            
	\mc{3}{l}{Number of unemployment spells in last 2 years (ref. 0 spells)} \\
\quad 1 spell &       0.032\sym{***}&      -0.015         &      -0.017\sym{**} &       0.002         &      -0.019\sym{**} &       0.002     \\

            &     \raisebox{.7ex}[0pt]{\scriptsize(0.010)}         &     \raisebox{.7ex}[0pt]{\scriptsize(0.009)}         &     \raisebox{.7ex}[0pt]{\scriptsize(0.008)}         &     \raisebox{.7ex}[0pt]{\scriptsize(0.003)}             &     \raisebox{.7ex}[0pt]{\scriptsize(0.008)}         &     \raisebox{.7ex}[0pt]{\scriptsize(0.003)}         \\

	\quad 2 spells&       0.011         &      -0.025\sym{***}&      -0.031\sym{***}&       0.005         &      -0.030\sym{***}&       0.006\sym{*} \\
	
            &     \raisebox{.7ex}[0pt]{\scriptsize(0.010)}         &     \raisebox{.7ex}[0pt]{\scriptsize(0.009)}         &     \raisebox{.7ex}[0pt]{\scriptsize(0.008)}         &     \raisebox{.7ex}[0pt]{\scriptsize(0.003)}            &     \raisebox{.7ex}[0pt]{\scriptsize(0.008)}         &     \raisebox{.7ex}[0pt]{\scriptsize(0.003)}         \\

\quad $\geq$ 3 spells&       0.011         &      -0.024\sym{**} &      -0.027\sym{***}&       0.003         &      -0.027\sym{***}&       0.003          \\

            &     \raisebox{.7ex}[0pt]{\scriptsize(0.011)}         &     \raisebox{.7ex}[0pt]{\scriptsize(0.010)}         &     \raisebox{.7ex}[0pt]{\scriptsize(0.008)}         &     \raisebox{.7ex}[0pt]{\scriptsize(0.004)}          &     \raisebox{.7ex}[0pt]{\scriptsize(0.008)}         &     \raisebox{.7ex}[0pt]{\scriptsize(0.004)}         \\

Last unemployment duration&      -0.001\sym{*}  &       0.001\sym{**} &       0.001         &       0.001\sym{***}&       0.001\sym{*}  &       0.001\sym{***}\\
            &     \raisebox{.7ex}[0pt]{\scriptsize(0.001)}         &     \raisebox{.7ex}[0pt]{\scriptsize(0.001)}         &     \raisebox{.7ex}[0pt]{\scriptsize(0.000)}         &     \raisebox{.7ex}[0pt]{\scriptsize(0.000)}               &     \raisebox{.7ex}[0pt]{\scriptsize(0.000)}         &     \raisebox{.7ex}[0pt]{\scriptsize(0.000)}         \\
\mc{7}{l}{\textbf{Personality traits}}\\

	Internal locus of control &                      &                     &                     &                     &       0.010\sym{***}&       0.001         \\
            &                     &                     &                     &                             &     \raisebox{.7ex}[0pt]{\scriptsize(0.003)}        &     \raisebox{.7ex}[0pt]{\scriptsize(0.001)}         \\

Conscientiousness&                     &                     &                     &                      &      -0.008\sym{**} &       0.000           \\
            &                     &                     &                     &                        &     \raisebox{.7ex}[0pt]{\scriptsize(0.003)}         &     \raisebox{.7ex}[0pt]{\scriptsize(0.002)}         \\

Openness&                     &                     &                     &                     &       0.017\sym{***}&       0.000      \\
            &                     &                     &                     &                       &     \raisebox{.7ex}[0pt]{\scriptsize(0.003)}         &    \raisebox{.7ex}[0pt]{\scriptsize(0.002)}         \\

Extraversion&                     &                     &                     &                       &       0.005\sym{*}  &       0.003\sym{**} \\
            &                     &                     &                     &                       &     \raisebox{.7ex}[0pt]{\scriptsize(0.003)}         &     \raisebox{.7ex}[0pt]{\scriptsize(0.001)}         \\
Neuroticism&                     &                     &                     &            &      -0.009\sym{***}&      -0.002            \\
            &                     &                     &                     &                      &     \raisebox{.7ex}[0pt]{\scriptsize(0.003)}         &     \raisebox{.7ex}[0pt]{\scriptsize(0.001)}         \\
[1em]
Objective benchmark $O_i$&                     &      -0.544\sym{***}&      -0.456\sym{***}&      -0.088\sym{***}&      -0.457\sym{***}&      -0.088\sym{***}\\
            &                     &     \raisebox{.7ex}[0pt]{\scriptsize(0.019)}         &     \raisebox{.7ex}[0pt]{\scriptsize(0.016)}         &     \raisebox{.7ex}[0pt]{\scriptsize(0.007)}         &                      \raisebox{.7ex}[0pt]{\scriptsize(0.016)}         &     \raisebox{.7ex}[0pt]{\scriptsize(0.007)}         \\
\hline \\[-2.0ex]
	No. of observations         &    5,376     &    5,376         &    5,376        &    5,376                &    5,200         &    5,200       \\
$R^2$         &       0.458         &       0.239         &       0.248         &       0.067         &       0.257         &       0.068       \\     	
Mean dep. variable & 7.183 &  0.170 & 0.205 & -0.035 & 0.205 & -0.036 \\
\\[-2.0ex]
\hline \hline
\end{tabular}
\end{footnotesize}
\begin{tablenotes} \scriptsize
\item \textit{Note:} The table reports the results of OLS regressions. In column (1), the dependent variable is individuals' subjective belief $S_i$ about their net monthly reemployment wage (in log). In column (2), we consider the log difference between the subjective belief $S_i$ and the objective benchmark $O_i$. In columns (3) and (5), we set negative values to zero and thus only exploit variation in positive deviations (``optimism''), while in columns (4) and (6), we set positive values to zero and thus only exploit variation in negative deviations (``pessimism''). Robust standard errors are shown in parentheses. ***/**/* indicate statistical significance at the 1\%/5\%/10\%-level.
\end{tablenotes}
\end{threeparttable}}
\end{table}

%Overall, we observe intuitive correlations between the accuracy of job seekers' beliefs and their background characteristics. For instance, 
It can be seen in column~(1) that women expect to earn about 17\% less than men upon reemployment \citep[see also][]{CLM2017,cortes2021gender,reuben2017preferences}. %This is in line with existing evidence that men generally have higher levels of self-confidence \citep{barber2001boys, } and set higher reservation wages \citep{CLM2017} than women. 
When we take into account the heterogeneity in job seekers' objective wage potential in column~(2), the gender gap remains negative, suggesting the difference between subjective expectations and objective benchmarks is about nine percentage points lower for women than for men. This suggests that about half of the gender gap in wage expectations is because men actually earn higher wages than women, while the other half is due to men being overly optimistic. We also observe that German citizens, who may possess a better understanding of the German labor market dynamics, have more accurate beliefs compared to foreigners. Moreover, high-skilled workers exhibit a greater tendency toward overoptimism compared to low-skilled workers. Quantitatively, job seekers with a university degree tend to overestimate their earnings potential by approximately 12 percentage points more compared to those without any higher education. This result is in line with earlier findings that higher levels of education are associated with individuals being more overconfident about their investment decisions \citep{bhandari2006demographics,trejos2019overconfidence}. 
%This pattern could reflect that individuals' beliefs about their abilities \citep{stinebrickner2012learning,wiswall2015determinants} or about the returns to schooling \citep{jensen2010perceived,attanasio2014education} affect  educational or occupational choices early on in their careers. %\footnote{\cite{Balleer2021} find that overconfidence among job seekers in the U.S. decreases with their skill level. In this context, one should note that our regressions condition on the level of objective predictions and various other covariates, enabling us to disentangle the effect of education from differential misperceptions along other characteristics that are correlated with job seekers' education.} 
Lastly, we find that wage optimism tends to be less pronounced among job seekers who have experienced more frequent periods of unemployment in the past. This observation aligns with the notion that individuals with greater job search experience may have accumulated more accurate information about their potential earnings.

% and it is therefore rational that men have also higher wage expectations. At the same time, the other half of the gender gap in wage expectations cannot be explained by the objective predictions suggesting that men tend to be more confident even in comparison to the rational benchmark. Considering the decomposition of results for optimistic and pessimistic beliefs in column~(3) and column~(4), we find that the gender differences are mainly driven by men exhibiting more overly optimistic beliefs rather than women being too pessimistic.

%Moreover, we observe several other noteworthy correlations. First, it may not be surprising that German citizens have more accurate beliefs compared to foreigners, as they may possess a better understanding of the German labor market dynamics. Second, 

In column~(5) and~(6), we also use survey information to explore the role of job seekers' personality traits in shaping belief inaccuracies. Notably, we find that job seekers who hold an internal locus of control, believing that their life outcomes are primarily determined by their own actions rather than external factors, tend to hold more optimistic beliefs about their future wages \citep[see also][]{CCU10,mcgee10}. At the same time, higher levels of openness and extraversion, as well as lower levels of neuroticism, are associated with greater optimism among job seekers. In this context, one should note that columns~(1) to~(4) of Table~\ref{tab:correlates} only include covariates used to generate the objective benchmarks, suggesting that the estimated correlations are not due to heterogeneity in rational expectations. In contrast, our prediction model does not incorporate the personality traits in columns (5) and (6). Thus, the observed patterns may reflect differences in workers' actual earnings potential rather than misperceptions. %If job seekers understand that they exhibit certain traits associated with higher earnings, it would be reasonable for them to adjust their wage expectations accordingly. 
To test if actual wage heterogeneity drives the results, Appendix Table~\ref{tab:realized_personality} shows the relation between personality traits and reemployment wages. Unsurprisingly, we find significant correlations between wages and most personality traits \citep[see also][]{heckman2006effects,mueller2006,heineck2010}. However, when controlling for objective benchmarks, these correlations mostly become insignificant, except for locus of control. This indicates that our predictions based on administrative covariates capture relevant heterogeneity among job seekers, suggesting that patterns in Table~\ref{tab:correlates} mainly reflect how personality traits relate to inaccuracies in wage beliefs.

\subsection{Anchoring of beliefs to pre-unemployment wages} \label{sec:anchor}

Various studies indicate that individuals often rely on anchoring heuristics \citep{KST1982} when forming their expectations. Related to our setting, it is commonly observed that unemployed job seekers anchor their reservation wages to their previous salary before becoming unemployed \citep[see, e.g.,][]{feldstein1984unemployment,KM2016,le2019unemployment,KMP2020}. However, in the absence of objective benchmarks in the previous literature, it is challenging to determine the extent to which this type of anchoring is justified. %In light of this, we now compare both job seekers' subjective wage expectations and the objective predictions to their pre-unemployment wages. 

\begin{figure}[h]
	\caption{Subjective beliefs and objective benchmarks relative to pre-unemployment wages \label{fig:pre_wage}}
	\centerline{
		\begin{threeparttable}
			\begin{footnotesize}
			\begin{tabular}{cc}
				\textbf{A. Distribution of changes} & \textbf{B. Relationship between changes} \\
				%compared to pre-unemployment wage & compared to pre-unemployment wage \\
				\includegraphics[width=0.45\linewidth]{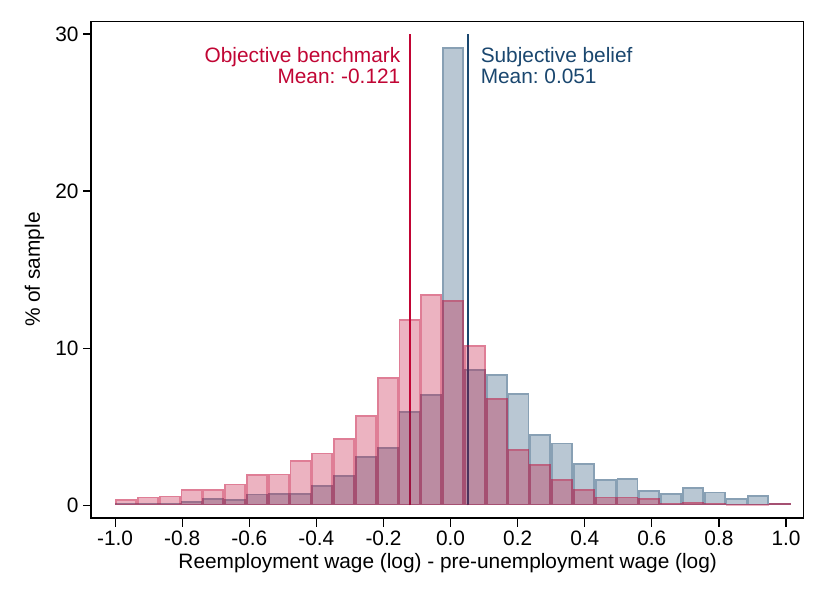} &
				\includegraphics[width=0.45\linewidth]{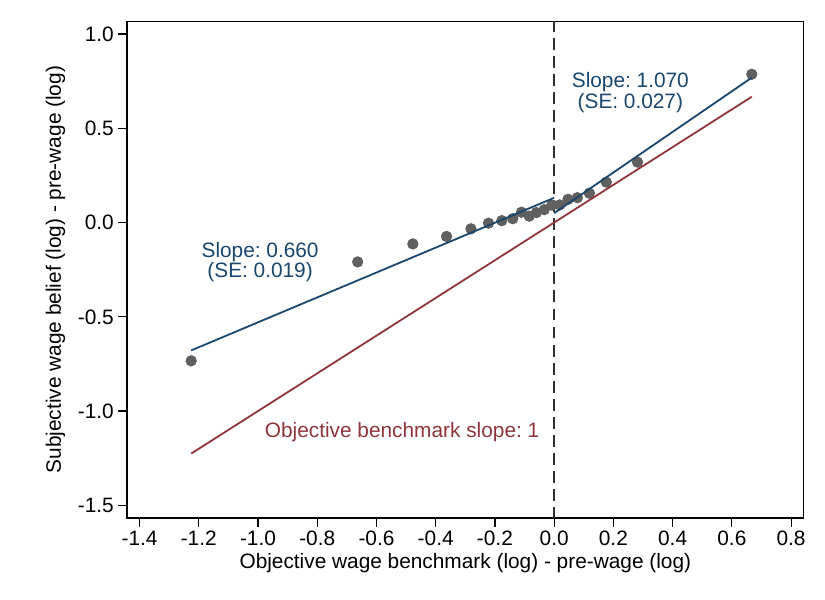}\\ 
			\end{tabular}
		\end{footnotesize}
			\begin{tablenotes} \scriptsize
				\item \textit{Note:}
				Panel~A shows a histogram of individuals' subjectively expected reemployment wage changes and objectively predicted reemployment wage changes (i.e. both in comparison to their pre-unemployment wages). Panel~B depicts a binned scatter plot (with 20 bins) for the individual-level relation between the two variables. Slope coefficients are estimated separately for positive and negative variation in objective wage changes. Both coefficients are significantly different from 1 (p-values $<$ 0.001 and $=$ 0.005, respectively). $N=5,376$.
				%The estimated slope is significantly different from 1 (p-value $<$ 0.001).
			\end{tablenotes}
	\end{threeparttable}}
\end{figure}

%			\includegraphics[width=0.7\linewidth]{graphs/p_subjective_objective_wagechange_survey_asymmetric}\\ 
%			\end{footnotesize}
%			\begin{tablenotes} \scriptsize
%				\item \textit{Note:}
%				The figure depicts a binned scatter plot (with 20 bins) for the relation between individuals' subjectively expected wage changes and objectively predicted wage changes (i.e. both in comparison to their pre-unemployment wages). Slope coefficients are estimated separately for positive and negative variation in objective wage changes. Both coefficients are significantly different from 1 (p-values  $<$ 0.001 and $=$ 0.005, respectively). $N=5,376$.

With this in mind, Panel~A of Figure~\ref{fig:pre_wage} shows the distributions of subjective beliefs and objective predictions relative to pre-unemployment wages. It turns out that job seekers overestimate their wage potential not only relative to the objective benchmarks generated from wages of similar workers, but also in comparison to their own previous salary. On average, job seekers anticipate a 5.1\% wage increase compared to their last job, whereas objective predictions indicate an average decrease of approximately 12.1\% in actual wages compared to the pre-unemployment wage. While periods of unemployment often come with wage penalties upon reemployment \citep{arulampalam2001unemployment,gregory2001unemployment}, it appears that job seekers, on average, do not account for these adverse effects of unemployment when forming their wage expectations. Moreover, the distribution of expected wage changes is much more compressed around zero compared to the distribution of changes in objective predictions. This suggests that job seekers anchor their wage expectations more strongly to their past wage than is objectively justified.

Panel~B of Figure~\ref{fig:pre_wage} further illustrates the relationship between job seekers' subjectively expected wage change and the objectively predicted wage change. Allowing for differential slopes for positive and negative variations in objective wage changes, reveals that the role of anchoring is asymmetric. Among individuals who are predicted to face a wage penalty compared to their previous wage, we estimate a slope coefficient of 0.66 (SE: 0.02). Being significantly smaller than the objective benchmark slope of one, this coefficient indicates that job seekers perceive their reemployment wage to be closer to their pre-unemployment wage than it actually is. When job seekers should reasonably anticipate a ten percentage point larger wage decline, they expect the wage decrease to be, on average, only 6.6 percentage points larger. That is, they tend to anchor their beliefs too strongly to their pre-unemployment salary. Conversely, job seekers who can reasonably expect a wage increase compared to their previous salary hold relatively accurate beliefs, as indicated by the slope coefficient of 1.07 (SE: 0.03). %Together, the results highlight that the role of anchoring in the belief formation process differs across job seekers with positive and negative outlooks for the future.}

\subsection{Belief updating over the unemployment spell} \label{sec:update}

Job seekers who acquire additional information during their search may revise their in response to the signals they receive \citep[see, e.g.,][for learning models of the labor market]{burdett1988declining,gonzalez2010equilibrium,conlon2018labor}. Therefore, we use wage expectations that were repeatedly elicited during the first and second survey waves to explore the evolution of job seekers' beliefs throughout their unemployment spell. We observe this repeated wage expectations for 459 individuals who were still unemployed and actively searching for a job at the time of the follow-up interview, conducted one year after entry into unemployment. Consistent with \cite{KM2016} and \cite{MuellerSpinnewijnTopa2021}, we observe that job seekers' overoptimism remains persistent throughout the unemployment spell and individuals update their wage expectations only to a very limited extent.\footnote{Appendix Figure~\ref{fig:wave2_hist} shows that there are no statistically significant changes in the distribution of wage expectations between the two waves ($p=0.872$ based on a Kolmogorov-Smirnov test for equal distributions in both waves).} 

\begin{figure}[h!]
	\caption{\label{fig:updating} Belief updating over the unemployment spell}
	\centerline{
		\begin{threeparttable}
			\begin{footnotesize}
				\begin{tabular}{cc}
%					\textbf{(a) Change in job finding beliefs} & \textbf{(b) Change in wage beliefs}\\
%					\includegraphics[width=0.45\linewidth]{graphs/p_wave2vs1_emp_change} &
%					\includegraphics[width=0.45\linewidth]{graphs/p_wave2vs1_wage_change}\\
%					\textbf{(c) Change in job finding overconfidence} & \textbf{(d) Change in wage overconfidence}\\ 
%					\includegraphics[width=0.45\linewidth]{graphs/p_wave2vs1_emp_bias} &
%					\includegraphics[width=0.45\linewidth]{graphs/p_wave2vs1_wage_bias} \\
					\textbf{A. Change in wage expectations}  & \textbf{B. Change in accuracy of wage expectations}\\ 
					\includegraphics[width=0.5\linewidth]{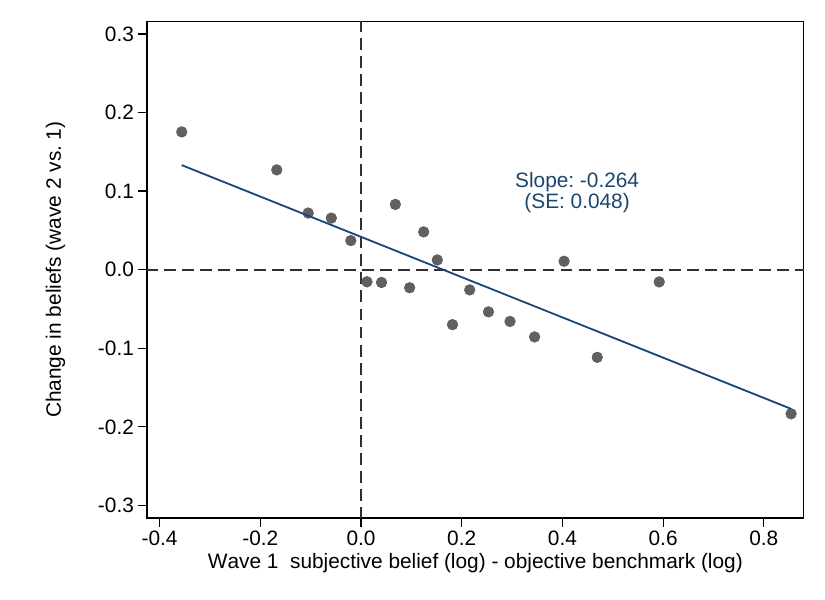} &				\includegraphics[width=0.5\linewidth]{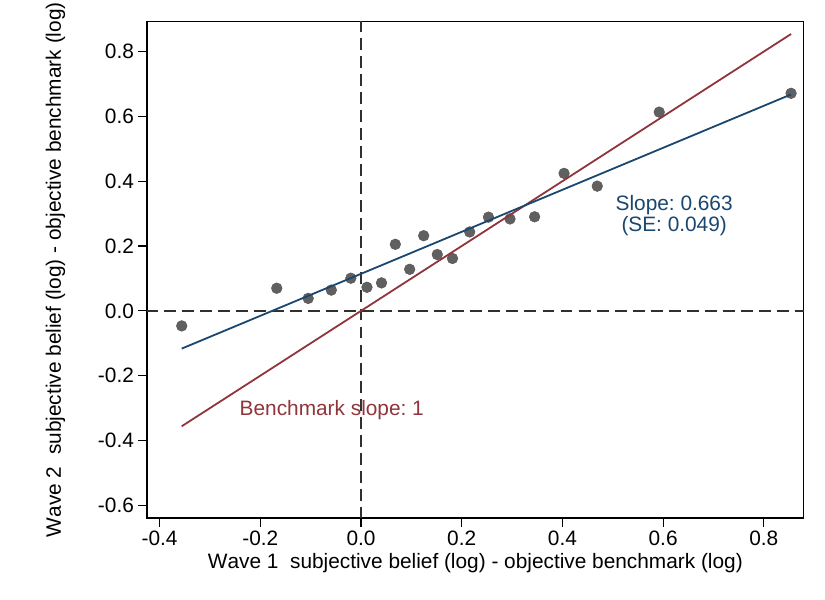} \\
				\end{tabular}
			\end{footnotesize}
			\begin{tablenotes} \scriptsize
				\item \textit{Note:}
				The figure depicts binned scatter plots of individuals' change in wage beliefs between waves 1 and 2 (Panel~A) and wage overconfidence in wave 2 (Panel~B) against the overconfidence in wave 1. Wave 1 was conducted 7 - 14 weeks after unemployment entry and wave 2 was collected 12 months after entry. The sample only includes individuals who are still in the same unemployment spell in wave 2. The estimated slope in Panel ~A is significantly different from 0 (p-value $<$ 0.001) and significantly different from -1 (p-value $<$ 0.001). The estimated slope in Panel ~B is significantly different from 0 (p-value $<$ 0.001) and significantly different from 1 (p-value $<$ 0.001). $N=459$.
			\end{tablenotes}
	\end{threeparttable}}
\end{figure}

Panel~A of Figure~\ref{fig:updating} illustrates the change in respondents' expectations from the first to the second interview against their initial wage optimism measured during the first interview. The estimated slope coefficient of -0.26 indicates that beliefs are not updated perfectly. Job seekers who initially overestimate their reemployment wage by an additional 10\% only decrease their wage expectations by 2.6\% more during the course of their unemployment spell.\footnote{It is worth noting that the estimated negative slope may be influenced to some extent by statistical mean reversion resulting from measurement error in belief elicitation, which suggests that the estimated slope coefficient might represent a lower bound for the true relationship.} Additionally, we observe that only job seekers who exhibit significant initial wage optimism in the first interview revise their wage expectations downwards. Conversely, job seekers who overestimate their wage potential by up to 17\% actually increase their wage expectations over time. These individuals most likely have received negative feedback from the jobs they encountered during their job search, yet they appear to resist updating their wage expectations accordingly.

Finally, we account for changes in objective benchmarks over time by predicting the reemployment wages of job seekers who remain unemployed for at least 12 months following their entry into unemployment, as observed in the administrative data. Consistent with previous evidence suggesting negative duration dependence \citep{kroft2013duration,eriksson2014employers,schmieder2016effect,zucht2023duration}, we find that the objectively predicted wage for respondents who remain unemployed decreases, on average, by about 6\% over the course of one year. Given that many job seekers are reluctant to revise their wage expectations downward, this implies that the majority of job seekers who remain unemployed display an even greater degree of overoptimism as time progresses. This is illustrated in Panel~B of Figure~\ref{fig:updating}, showing the relationship between the deviations of subjective beliefs and objective benchmarks during the first and second interview waves. The estimates suggest that the degree of wage optimism increases for job seekers who initially overestimate their wage potential by up to 34\% (i.e. the intersection of the red and blue lines in Panel~B of Figure~\ref{fig:updating}) and only decreases for those who exhibit an even higher level of optimism during the first interview.

It is important to note that the group of individuals who are still unemployed during the second interview after one year consists of job seekers with poor overall labor market prospects (reflecting dynamic selection over the unemployment spell). The reluctance to revise their wage expectations might be one factor that hinders the labor market integration of these job seekers, contributing to their extended unemployment spell. Conversely, job seekers who have adjusted their wage expectations may have secured employment before their second interview, making it difficult to draw conclusions as to how their beliefs were updated over time.

\subsection{Perceived control over reemployment wages} \label{sec:incentives}

Up to this point, we have examined job seekers' beliefs while treating their job search behavior as given. However, another important dimension involves job seekers' beliefs about how changes in their search behavior can affect reemployment wages.\footnote{\cite{Spinnewijn2015} introduces this distinction in the context of job seekers' beliefs about their job finding prospects. An individuals' baseline belief describes the expected job finding probability for a given search strategy, while the control belief refers to the change in the job finding probability when searching more intensively.} For example, by submitting a greater number of applications, job seekers enhance their prospects of attracting offers that come with particularly high wages. Therefore, job seekers may have some degree of control over their reemployment wages.
 %Moreover, this effect might be reinforced in the presence of negative duration dependence, e.g., due to skill depreciation during the unemployment spell. Job seekers who anticipate finding a job after a shorter period of unemployment due to their intensified search efforts may also anticipate receiving more favorable job offers, since their skills are perceived to have depreciated at a lesser rate \citep[see, e.g.,][]{nekoei2017does}. 	

In what follows, we examine how individuals' search behavior influences their subjective wage expectations, as well as their realized reemployment wages. This exercise comes with a non-trivial identification problem, because an individual's beliefs may not only depend on her choices, but her choices may also depend on her beliefs \citep{mu/sp/2022}. To address this concern, we explore exogenous variation in the incentives of unemployed workers to search for jobs and analyze their impact on job seekers' wage optimism. We exploit regional variation in the risk that job seekers will be subject to punitive sanctions along the administrative borders of local employment agency (LEA) districts. These sanctions involve temporary reductions in unemployment benefit payments and are imposed by caseworkers when they detect that job seekers are not complying with job search requirements.\footnote{Instances of non-compliance include insufficient job applications, rejecting job offers from the employment agency, and voluntary termination of employment.} The regional variation arises because LEAs have autonomy in deciding about the local policy style \citep[see, e.g.,][]{FSS2006,boockmann2014intensifying,DoerrKruppe2015}, including how strictly they punish job seekers for inadequate search behavior.

%role of anchoring heuristics and belief updating as potential drivers of job seekers' wage optimism. Another source may be found in the beliefs that job seekers hold about the consequences of their own behavior. Following Spinnewijn (2015), we can differentiate between \textit{baseline} and \textit{control} beliefs. The former captures the baseline expected wage, given effort the job seekers puts in, while the latter denotes the expected increase in wages when searching more. A job seeker is said to be control-optimistic when she overestimates the effect of her search effort on reemployment wages, i.e., when $\frac{\partial \text{Expected wage}}{\partial \text{Effort}} > \frac{\partial \text{Realized wage}}{\partial \text{Effort}}$. Importantly, the control beliefs is what should impact job seekers' search behavior. A control-optimist is expected to exert more effort as she expects larger returns. To shed light on job seekers' control beliefs, in the following, we seek to examine how individuals' search effort influences their expected and .}  

As a result, caseworkers in LEA districts with higher sanction intensities may exert greater pressure on job seekers, leading them to perceive stronger incentives to apply for and accept jobs. This can yield different implications for job seekers' wage expectations. On the one hand, sending out more applications may increase the likelihood of receiving high-wage offers, leading job seekers to adapt more optimistic beliefs about their reemployment wages.  On the other hand, the increased sanction risk may cause job seekers to become less selective, resulting in lower wage expectations.

\paragraph{Econometric strategy:} Our econometric strategy in this section explores regional differences in job seekers' risk of facing benefit sanctions. We leverage
data on the annual number of benefit sanctions imposed in the year before each job seeker entered unemployment, normalized by the stock of unemployed workers in each of the 178 LEA districts. We incorporate the resulting sanction intensity, $SI_{j}$, into a border-pair fixed-effects models of the following form \citep[see also][]{dube2010minimum,Caliendo2022}:
\begin{align}
	Y_{ijb} &= \alpha + \delta SI_{j} + \beta X_{i} + \phi R_{j} + \kappa_b + \varepsilon_{ijb}, \label{est:base}
\end{align}
where $i$ denotes the individual job seeker, $j$ the LEA district in which the individual is located at the beginning of the unemployment spell, and $b$ a pair of bordering LEA districts such that $\kappa_b$ denotes the border-pair fixed effects for any combination of two neighboring LEA districts. Moreover, $R_j$ captures regional characteristics including the local unemployment rate, vacancy rate, gross domestic product, industry structure, and federal state fixed effects, and $X_i$ accounts for individual-level characteristics. Standard errors are clustered at the LEA district level. The parameter of interest $\delta$ identifies the effect of sanction intensity on the outcome variables $Y$ by comparing individuals living in similar, neighboring LEA districts but facing varying risks of being sanctioned. In Appendix \ref{app:sanc}, we present additional details about our econometric strategy and provide empirical evidence supporting the validity of the identifying assumptions. Specifically, we demonstrate (1) that bordering LEA districts exhibit similar local labor market conditions, (2) that the sanction intensity is independent of regional indicators and job seekers' individual characteristics and (3) that other dimensions of regional policy styles do not systematically vary with local sanction intensities.

%\AS{
%	\bi
%	\item Should we already mention here that we don't find effects on reservation wages and can therefore interpret the effects as perceived returns to search effort?
%	\item In line with \citet{Schneider2008} who finds no effect of sanctions on the reservation wage for UB II recipients
%	\ei
%}
%Our model assumes that job seekers accept every wage offer, thereby assuming that any reservation wage is not binding. This central assumption is motivated by recent empirical evidence suggesting that reservation wages are not the main driver of search dynamics and the job finding hazard (e.g. Card et al., 2007; Krueger and Mueller, 2016; Schmieder et al., 2016).
%But descriptive results on job finding suggest that reservation wages do matter: otherwise we would not see lower job finding among the most overconfident individuals

\paragraph{Effect of sanction risk on behavior and beliefs:} Table~\ref{tab:sanction} shows the effect of the sanction risk on job seekers' search effort, wage expectations, and realized wages. In line with standard search-theoretical arguments, a stricter sanction regime seems to motivate unemployed workers to exert more search effort. Specifically, as shown in column~(1) of Table~\ref{tab:sanction}, a ten percentage point higher sanction risk -- equivalent to an increase of approximately one standard deviation -- raises the number of weekly job applications by about 9.8\% ($p=0.016$). At the same time, the estimates in column~(2) reveal that a stricter sanction regime fosters greater optimism among job seekers regarding their earnings potential. Raising the sanction risk by ten percentage points increases job seekers' wage expectations relative to the objective benchmark by about 1.8\% ($p=0.006$). When we differentiate between individuals who overestimate and underestimate the potential wages they could earn, we observe that the sanction risk impacts both dimensions. Specifically, it significantly enhances optimism (as indicated in column~(3)) while concurrently reducing pessimism (as shown in column~(4)). 

\begin{table}[h!]
	\centerline{
		\begin{threeparttable}
			\caption{Effect of sanction risk on search behavior and accuracy of wage expectations}\label{tab:sanction}
			\tabcolsep=0.2cm
			\begin{footnotesize}
				\begin{tabular}{lccccccc}
					\hline\hline \\[-2.0ex]
				\\[-2.0ex]
				& && \mc{3}{c}{ Accuracy of wage expectations$^{(b)}$}\\
	\\[-2.0ex]
\cline{4-6}
\\[-2.0ex]
Dependent variable	&  Log no. of job &&  $S_i - O_i$ & $S_i - O_i$ & $S_i - O_i$ &&  Log realized net  \\
			& applications$^{(a)}$   &   &	  & Pos. values & Neg. values && monthly wage$^{(c)}$ \\
			& (1) && (2) & (3) & (4) && (5)  \\
\\[-2.0ex]
\hline
\\[-2.0ex]					
Effect of sanction intensity  &      0.098$^{**}$ & & 0.018$^{***}$&      0.011$^{**}$&      0.007$^{***}$ &&  -0.009 \\
	&      \raisebox{.7ex}[0pt]{\scriptsize(0.040)}     &&      \raisebox{.7ex}[0pt]{\scriptsize(0.007)} &      \raisebox{.7ex}[0pt]{\scriptsize(0.005)} &      \raisebox{.7ex}[0pt]{\scriptsize(0.003)} &&  \raisebox{.7ex}[0pt]{\scriptsize(0.008)}\\
No. of observations         						&       5,669  &&  5,669 &   5,669 &  5,669 && 17,973 \\
Mean dep. variable   							&       1.716  	&&  0.125 &   0.167&     -0.042 &&   7.001 	  \\
\\[-2.0ex]
\hline \hline
\end{tabular}
\end{footnotesize}
\begin{tablenotes} \scriptsize 
\item \textit{Note:} The table reports the effect of the local sanction intensity (measured in 10\%-points) on job seekers' search effort, their subjective wage expectations, and realized wages.  In all specification, we account for socio-demographic and regional characteristics, as well as border-pair fixed effects. Standard errors clustered at the LEA district level are shown in parentheses. ***/**/* indicate statistical significance at the 1\%/5\%/10\%-level. 
\item $^{(a)}$ In column (1), the dependent variable is the log number of job applications sent since unemployment entry.
\item $^{(b)}$ In column (2), the dependent variable is the log difference between subjective belief $S_i$ and objective benchmark $O_i$. In column (3), we set negative values to zero and thus only exploit variation in positive deviations (``optimism''), while in column (4), we set positive values to zero and thus only exploit variation in negative deviations (``pessimism'').
\item $^{(c)}$ In column (5), the dependent variable is the log realized wage of individuals observed in the administrative sample who start a regular job within 24 months after entry into unemployment.
\end{tablenotes}
\end{threeparttable}}
\end{table}

%\AS{
%	\bi
%	\item Add effects on reservation wages to table and in interpretation
%	\item Focus interpretation on "perceived returns to search effort" by comparing effects on expected vs. realized wages
%	\ei
%}

These findings suggest that job seekers perceive to have control over their reemployment wages, possibly because they believe sending out a larger number of applications increases their chances of attracting high-wage offers.\footnote{This effect might be reinforced in the presence of negative duration dependence, e.g., due to skill depreciation during the unemployment spell. Job seekers who anticipate finding a job after a shorter period of unemployment due to their intensified search efforts may also anticipate receiving more favorable job offers, since their skills are perceived to have depreciated at a lesser rate \citep[see, e.g.,][]{nekoei2017does}.} Moreover, the positive effect on wage optimism indicates that this control belief outweighs the potential direct effect of the sanction risk that would lead job seekers to become less selective. Yet, the increased optimism does not translate into higher realized wages. Instead, as shown in column~(5) of Table~\ref{tab:sanction}, realized wages upon reemployment tend to be (insignificantly) lower when job seekers are subject to a more restrictive sanction regime. A plausible explanation for these patterns is that job seekers not only exhibit overly optimistic beliefs about their reemployment wages for a given search strategy but also overestimate the influence of their own behavior.

\section{Labor Market Implications}\label{sec:discussion_implications}

Misperceptions about labor market prospects can distort job seekers' decision-making, potentially prolonging unemployment. For example, \cite{mu/sp/2022} suggest overly optimistic beliefs about wage offers as a cause of job seekers underestimating their risk of long-term unemployment because it can lead them to be excessively selective in accepting job offers. Against this backdrop, a natural policy response would be to provide job seekers with information about their objective earnings potential in order to correct their overly optimistic beliefs. This could potentially reduce job seekers' selectivity, thereby increasing their chances of reemployment. Such a policy can be effective if optimistic beliefs are merely a consequence of incomplete information about the labor market. Supporting this idea, we observe that job seekers with greater unemployment experience tend to hold more accurate earnings expectations. Moreover, the anchoring of beliefs to pre-unemployment wages may stem from workers' having incomplete information, leading them to utilize their previous wage as a signal for their wage upon reemployment.

However, our findings also indicate that information policies can lead to unintended consequences. Specifically, wage optimism is most noticeable among individuals at the lower end of the objective benchmark distribution and among those predicted to face a wage penalty relative to their previous salary. Correcting these misperceptions may discourage individuals, potentially reducing their effort or causing them to abandon their job search altogether. For workers with low objective benchmarks, the prospect of finding a job with a wage that matches their actual earnings potential may not provide a sufficient incentive to motivate them to search.\footnote{Aligning with this idea, experimental evidence from Uganda by \cite{bandiera2023search} indicates that an intervention matching workers with potential employers, which reduces their optimistic beliefs about labor market prospects, discourages job seekers, causing them to search less intensively for jobs. In a similar vein, \cite{banerjee2023learning} document discouragement among job seekers in South Africa who revise their beliefs in response to transport subsidies that reduce their search costs.}
	
 %These concerns are particularly serious because many job seekers in Germany have an objective earnings potential that is close to their monthly unemployment benefit level. \textbf{RM: Add some facts/figure about relation between benefits and benchmarks/beliefs} For instance...

At the same time, it is unclear to what extent job seekers would internalize the information they receive. It is often argued that subjective beliefs serve essential psychological needs \citep[see, e.g.,][for an overview]{benabou2016mindful}. For instance, individuals may hold unrealistically positive beliefs because they derive direct utility from maintaining a positive self-image \citep{brunnermeier2005optimal, koszegi2006ego} or they use optimism to enhance their motivation and overcome self-control problems \citep{Be/Ti/2002,Be/Ti/2004}. Consistent with these ideas, our findings indicate higher levels of overoptimism among job seekers with a lower objective earnings potential. This specific group of unemployed workers may have a heightened desire for motivated beliefs compared to individuals who can reasonably anticipate higher wages. The presence of motivated beliefs may pose distinct challenges, as they could induce individuals to deliberately suppress negative feedback to maintain their biased perceptions \citep{Be/Ti/2002,Be/Ti/2004,zimmermann2020dynamics}. Aligning with this notion, we find that job seekers who remain unemployed for an extended period are reluctant to revise their wage expectations downward, even though they have most likely received negative signals during their job search. We also have shown that job seekers who perceive greater extrinsic incentives to apply for jobs maintain more optimistic wage expectations, despite not experiencing an actual increase in wages. It is plausible that job seekers facing increased pressure from their caseworkers to submit more job applications adjust their wage expectations to enhance their motivation to exert more effort.

In the following, we present descriptive empirical evidence on the potential labor market implications by analyzing the relationship between job seekers' wage optimism, their search behavior, and realized labor market outcomes. Before doing so, we first outline the potential mechanisms at play within a job search framework in which unemployed workers decide on their search strategies while facing uncertainty about potential wages.

\subsection{Theoretical considerations} \label{sec:implications_theory}
%To illustrate the potential labor market implications of overly optimistic beliefs, we present a job search framework where unemployed workers decide on their search strategies while facing uncertainty about potential wages.

%\paragraph{Job search framework:} 
While searching for jobs individuals receive a flow utility of $b$, capturing the value of UI benefits net of any utility costs associated with maintaining eligibility. They decide about the number of job applications they send out, $s$, and their reservation wage, $\phi$, the minimal wage offer they would accept. The probability of a successful job application, resulting in a job offer, is denoted by $\lambda$, while the effort costs incurred during the job search are captured by the increasing and convex function $\gamma(s)$. Each job offer is associated with a wage, denoted by $w$, which is a random draw from an exogenous wage offer distribution $F(w) \sim N(\mu, \sigma)$. If the wage offer exceeds her reservation wage, the job seeker will accept the job and receive a wage of $w$ for the rest of time. If not, she will reject the offer and continue searching. Upon receiving multiple job offers, individuals accept the highest wage offer $y=\max\{w_1,w_2,...,w_n\}$ if it exceeds their reservation wage. The distribution of this maximum offer can be described as $F_{y}(y; \mu, \sigma)=F(y; \mu, \sigma)^{n}$, where $n=\lambda s$ represents the total number of job offers received.

Inspired by \cite{cortes2021gender}, job seekers hold subjective beliefs about the mean of wage offer distribution $\hat{\mu}$ and maximize their utility as if $\hat{\mu}$ were the true mean. Similar to \cite{cortes2021gender}, we assume that job seekers are myopic, that is, they do not take into account potential future learning resulting from their current choices.\footnote{This assumption is supported by the absence of belief updating during the unemployment spell, shown in Section~\ref{sec:update}.}  When choosing their search strategy, individuals maximize their perceived present value of income:
\begin{align}
	U= \max_{s,\phi} b - \gamma(s) + \rho \left\{U  + \left[1 -(1-\lambda)^{s}\right] \int_{\phi}^{\infty} \left(V(y) - U \right)dF_{y}(y; \widehat{\mu}, \sigma)\right\} \label{eq:max}
\end{align}
where $\rho$ denotes the discount factor and $V$ describes the value of being employed at wage $y$. In this framework, individuals' optimal decisions have a simple characterization. The reservation wage $\phi$, for a given number of applications $s$, is set such that the job seeker is indifferent between accepting a job offer and remaining unemployed, meaning $V(\phi)=U$. The optimal number of application $s^*$, trading off the cost of search and its perceived returns, follows from the first-order condition of Equation~\ref{eq:max}.

%\paragraph{Consequences of wage optimism:} 
Job seekers are considered to be optimistic about their reemployment wages if $\hat{\mu} > \mu$. This optimism may affect their labor market integration by influencing their decisions about the reservation wage and the number of applications they submit. First, it is straightforward that the reservation wage increases in the degree of optimism, $\frac{\pa \phi}{\pa \widehat{\mu}}>0$. Optimistic job seekers anticipate higher future wage offers, which makes it more attractive to turn down a given offer and continue searching. Second, wage optimism may affect the optimal number of job applications. Specifically, optimistic beliefs enhance motivation to search, because individuals who anticipate higher wage offers perceive greater returns from each application, $\frac{\pa s^*}{\pa \widehat{\mu}} >0$.\footnote{It follows directly from the first-order condition of Equation~\ref{eq:max} that more optimistic wage expectations increase the optimal effort level for a given reservation wage. Simultaneously, job seekers adjust their reservation wages, which induces an income effect. This effect leads optimistic job seekers to reduce their search effort, as they prefer to wait for a better opportunity rather than quickly settling for a lower-paying job. However, under standard functional form assumptions (i.e. convex search costs), the motivating effect dominates, resulting in $\frac{\pa s^*}{\pa \widehat{\mu}} >0$.} %This motivational effect of optimistic beliefs is particularly pronounced when the marginal costs of effort are low or when the chances of success from additional applications are high.

%increases the optimal number of applications, , because individuals who anticipate higher wage offers perceive greater returns from each application. Assuming that job seekers experience diminishing returns to higher wages (i.e. $V'(y)> 0$ and $V''(y)<0$), the motivational effect of optimistic wage expectations is particularly pronounced for individuals with low objective earnings potential.

The expected job finding probability is given by: $\left[1 -(1-\lambda)^{s}\right](1-F(\phi; \widehat{\mu}, \sigma))$. Combining the effects on the reservation wage and search effort, the impact of wage optimism on job finding is ambiguous. On the one hand, optimistic beliefs encourage job seekers to be more selective, which lowers their job finding prospects. On the other hand, wage optimism can motivate unemployed workers to exert more effort, thereby increasing their job finding prospects. At the same time, wage optimism unambiguously leads to higher realized wages, as both mechanisms indicate a positive relationship between optimistic beliefs and actual wage outcomes.

\subsection{Descriptive evidence} \label{sec:implications_results}

%In the final part of our analysis, we take a closer look at the potential labor market implications of wage optimism. As elaborated upon in Section~\ref{sec:theory}, inaccurate beliefs can yield different consequences for the labor market integration of unemployed workers. On the one hand, optimistic wage expectations can motivate job seekers to exert more effort, thereby facilitating job finding. On the other hand, unemployed individuals who possess unrealistically optimistic wage expectations might exhibit an excessive degree of selectivity and reject job offers more frequently than justified. This in turn may lead to higher realized wages, but may prolong unemployment and cause job seekers to overestimate their reemployment prospects. 

In a last step of our analysis, we present descriptive evidence illustrating the empirical significance of the mechanisms outlined in the previous subsection. It is important to note that individuals' search behavior and their beliefs are closely intertwined, making it notoriously challenging to identify the causal effects of optimistic wage expectations on individuals' job search behavior and labor market outcomes. Nonetheless, as we shall see, the data patterns reveal insightful perspectives regarding the importance of the mechanisms at play.

%\textbf{In order to shed more light on this trade-off, as a last step of our analysis, we seek to examine the labor market implications of wage overoptimism. Empirically studying the consequences of beliefs for job seekers' search behavior and realized outcomes is notoriously challenging. From our stylized job search model in Section~\ref{sec:theory} it becomes clear that individuals' search behavior and their beliefs are closely intertwined. Not only are the choices made by job seekers influenced by their beliefs, their beliefs are also a function of their choices. In line with that, in Section~\ref{sec:incentives} we provided evidence that job seekers who engage in more intensified search activities due to external incentives tend to demonstrate higher levels of overoptimism. Moreover, our findings that point towards overoptimism being motivated also underline the endogeneity of beliefs. If optimistic expectations are more prevalent among individuals with higher self-image concerns or those who seek to overcome self-control problems from present-biased preferences, these factors may also confound the observed relation between overoptimism and future outcomes.}

\begin{figure}[h!]
	\caption{Descriptive evidence on labor market implications}  \label{fig:bias}
	\centerline{
		\begin{threeparttable}
			\begin{footnotesize}
				\begin{tabular}{cc}
					\textbf{A. Log no. of job applications$^{(a)}$} & \textbf{B. Log realized net monthly wage$^{(b)}$}\\
					\includegraphics[width=0.5\linewidth]{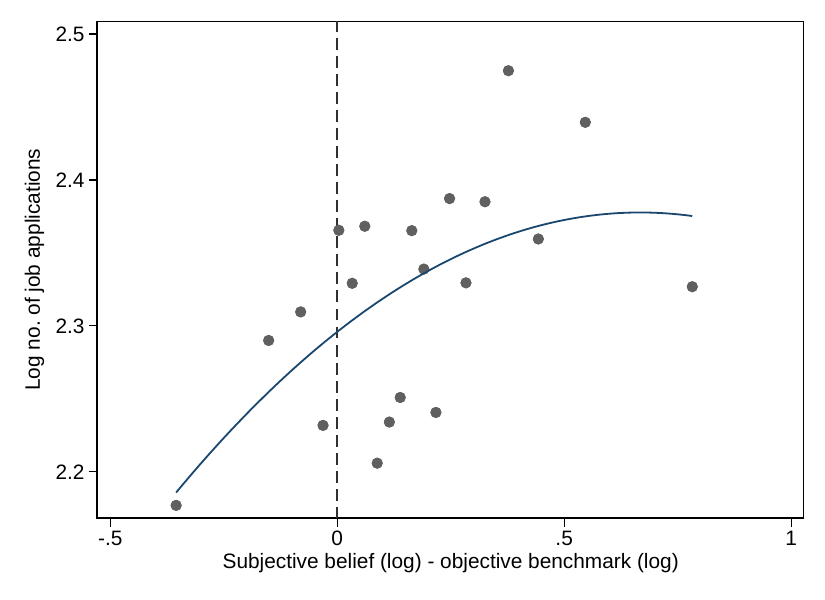} &\includegraphics[width=0.5\linewidth]{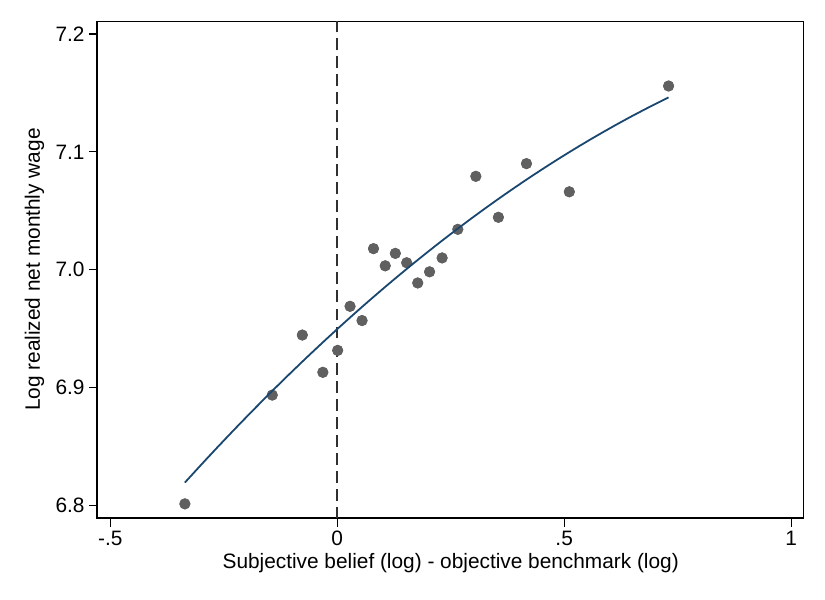} \\ 
					\textbf{C. Perceived job finding rate$^{(c)}$} & \textbf{D. Realized job finding rate$^{(d)}$}\\
					\includegraphics[width=0.5\linewidth]{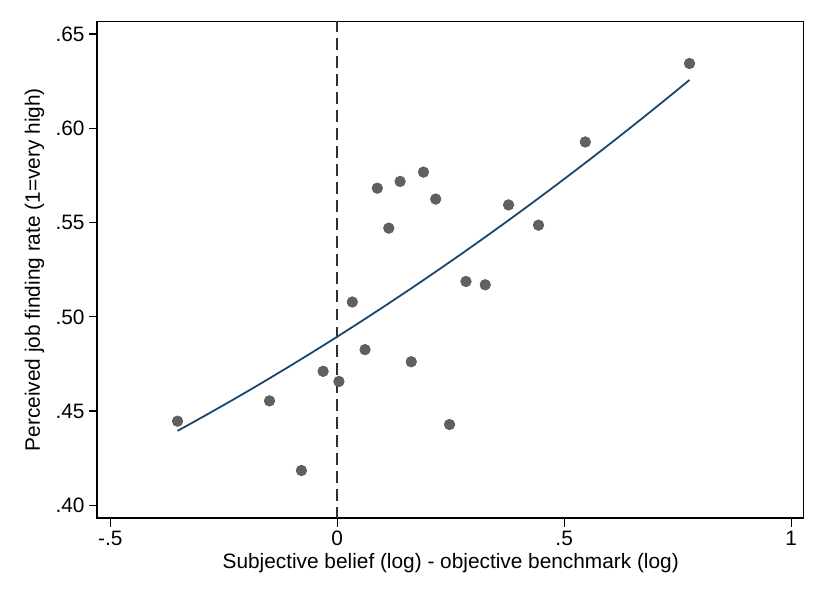}  
					&\includegraphics[width=0.5\linewidth]{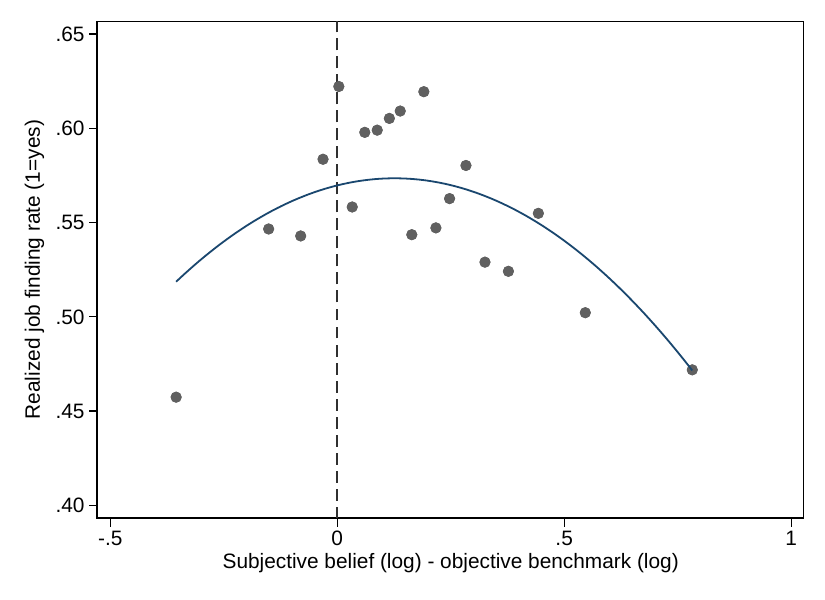} 
				\end{tabular}
			\end{footnotesize}
			\begin{tablenotes} \scriptsize
				\item \textit{Note:}
				The figure shows binned scatter plots (with 20 bins) of individuals' search effort, realized reemployment wage, and perceived and realized six-months-ahead job finding rates against their wage belief inaccuracy (defined as the log difference between subjective belief $S_i$ and objective benchmark $O_i$). The blue line shows a quadratic fit for the conditional means after residualizing against socio-demographic characteristics (gender, age categories, German citizen, education categories, married, any children, East Germany) and the objective benchmark. The corresponding regression results are reported in Appendix Table~\ref{tab:bias}.
				\item $^{(a)}$ In panel A, the dependent variable is the log number of job applications sent since unemployment entry.
				\item $^{(b)}$ In panel B, the dependent variable is the first monthly net wage of individuals who start a regular job within 24 months after entry into unemployment.
				\item $^{(c)}$ In panel C, the dependent variable is an indicator of whether the individual reports the perceived six-month-ahead job finding rate to be `very high'.
				\item $^{(d)}$ In panel D, the dependent variable is an indicator of whether the individual actually starts a regular job within the next six months.
			\end{tablenotes}
	\end{threeparttable}}
\end{figure}

Figure~\ref{fig:bias} illustrates correlations of job seekers' wage optimism or pessimism with their search effort, their realized wages, as well as their perceived and realized job finding rates. %\textbf{Jason: Binscatter is problematic in that it imposes linearity assumptions, I think a better plotting method, especially for non-linearity, is binsreg. I got this comment from a JoLE referee for my inside scoop paper.} 
Specifically, we depict binned scatter plots conditioning on socio-demographic characteristics and objective benchmarks (Appendix Table~\ref{tab:bias} reports the corresponding regression results). We find that the level of wage optimism is positively related to the number of job applications (see Panel~A of Figure~\ref{fig:bias}). In particular, job seekers who overestimate their wage by an additional 10\% sent out, on average, 1.6\% more job applications ($p=0.001$). Moreover, it turns out that workers who exhibit a greater level of optimism also earn higher wages upon finding a job within two years after the start of the unemployment spell (see Panel~B of Figure~\ref{fig:bias}). A wage expectation that is inflated by 10\% comes along with approximately 3.1\% higher monthly wages ($p<0.001$) in comparison to individuals who provide an accurate assessment of their earnings potential.

We also examine the relationship between the accuracy of job seekers' wage expectations and their perceived and actual job finding rates, both measured over a period of six months following the interview. This comparison reveals a remarkable pattern. On the one hand, as illustrated in Panel~C of Figure~\ref{fig:bias}, there is a positive and almost linear relationship between wage optimism and the perceived job finding probability. This suggests that job seekers who are most optimistic about their reemployment wages also report the highest perceived chances of finding a job. On the other hand, Panel~D  of Figure~\ref{fig:bias} shows a non-linear connection between the accuracy of individuals' wage expectations and their actual job finding rates. Specifically, job seekers who hold relatively accurate beliefs about their reemployment wage have the highest likelihood of finding a job within six months, whereas those who over- or underestimate their earnings potential face reduced reemployment rates. Consequently, our results suggest that the more optimistic workers are about the wages they can earn upon reemployment, the higher the likelihood that they will overestimate their prospects of finding a job. 

Taken together, the observed pattern aligns with the notion that optimistic job seekers, upon receiving wage offers they deem to be ``too low'', exhibit an excessive degree of selectivity. While optimist individuals may earn higher wages, they prolong the duration of unemployment beyond their initial expectations, resulting in a wedge between the true and perceived job finding rates. Concurrently, wage optimism can serve as a motivation to search more intensively, potentially explaining the positive correlation between optimistic beliefs and search effort. However, it appears that the impact of heightened effort on job finding is outweighed by the increased selectivity linked to job seekers' excessive optimism.

\section{Conclusion} \label{sec:concl}

%Job seekers' misperceptions about the labor market can distort their decision-making and prolong unemployment. 

In our study, we use objective benchmarks for the subjective wage expectations of unemployed workers, which 
enables us to document significant variation in the levels of optimism and pessimism among different groups of job seekers. Most notably, we find that workers with the lowest objective earnings potential and those expected to experience a wage decline compared to their previous job tend to exhibit the highest degree of overoptimism regarding their reemployment wages. Upon further investigating the nature of these optimistic beliefs, we observe that they persist throughout the unemployment spell and that job seekers believe to have control over their reemployment wage.

Finally, we have established a connection between overly optimistic wage expectations and job seekers' tendency to overestimate their prospects of reemployment. This finding holds important implications for policymakers committed to preventing long-term unemployment. To be more precise, our results suggest a wedge between the perceived and actual job finding rates for increasing levels of wage optimism. This aligns with the idea that inaccurate beliefs may incur decision-making costs, as overly optimistic job seekers may prolong unemployment by being excessively selective with respect to the jobs they accept. Against this backdrop, a natural policy response would encourage job seekers to revise their overly ambitious aspirations, which can enhance the reemployment prospects of unemployed workers. However, policymakers should be aware that information policies can have unintended consequences since optimistic beliefs may enhance job seekers' motivation to search.

%could present an attractive path for labor market policy focused on enhancing 

While the combination of survey data and administrative records offers valuable insights into the interrelation of individuals' beliefs, their job search behavior, and their actual labor market outcomes, our setting does not come without limitations. For instance, individuals may possess private information about their earnings potential that is not accounted for in our benchmarks. This poses a significant challenge in identifying misperceptions at the individual level, and has the potential to influence the relationship between subjective beliefs and job search or labor market outcomes. The ideal survey should elicit individuals' beliefs not only about their own earnings potential, but also about primitives, such as the perceived returns to search and the distribution of wage offers, to reduce issues of reverse causality. %Moreover, one could hope to improve the accuracy of objective benchmarks by conditioning on a richer set of commonly unobserved individual characteristics, including personality traits, non-cognitive skills, or preferences over non-wage job characteristics. %\textbf{XXX Add more stuff? XXX.}

Acknowledging these limitations, we consider our analysis as an important step toward opening the black box of how job seekers form their beliefs and how beliefs can affect their decisions while searching for jobs. Notably, various pieces of evidence substantiate the idea that motivated beliefs have a significant influence on job seekers' tendency to be overly optimistic about their labor market prospects. At the same time, information frictions might be at work as well, and untangling these distinct explanations presents an intriguing avenue for future research. For instance, it would be particularly interesting to provide a randomly selected group of job seekers with information about their objective earnings potential and analyze the consequences for their behavior and reemployment prospects. Such an approach would help to assess the significance of information frictions in influencing job seekers' decision-making and outcomes during the job search process. At the same time, analyzing how job seekers recall the provided information and update their beliefs depending on their priors could offer further insights into the role of motivated beliefs.

\newpage\clearpage
\renewcommand{\baselinestretch}{1.12}
\setlength{\baselineskip}{9pt}
\phantomsection
\addcontentsline{toc}{section}{References}
\bibliography{litbank}

% ---------------------------------------------------------------------
%  FIGURES AND TABLES
% ---------------------------------------------------------------------
\clearpage 
\renewcommand{\baselinestretch}{1.5}
\setlength{\baselineskip}{20pt}

%\section*{Figures and Tables}

%\input{tables/determinants_optimism_v4}

\newpage\clearpage

\begin{appendix}
	
	\bigskip
	\begin{center}
		\section*{{\LARGE Online Appendix}}
	\end{center}

\startcontents[sections]
\printcontents[sections]{l}{1}{\setcounter{tocdepth}{2}}

\newpage 

\renewcommand{\thesection}{A}
\renewcommand{\thetable}{A.\arabic{table}}
\setcounter{table}{0}
\renewcommand{\thefigure}{A.\arabic{figure}}
\setcounter{figure}{0}

\section{Additional Figures and Tables}

\begin{figure}[h!]
	\caption{\label{fig:robustness} Robustness: relationship between subjective beliefs and objective benchmark}
	\centerline{
		\begin{threeparttable}
			\begin{footnotesize}
				\begin{tabular}{ccc}
					%	\textbf{A. Baseline}  & \textbf{B. Restricted to reemployed after 24 months}&\textbf{C. Predition from alternative training data} \\ 
					%	\includegraphics[width=0.4\linewidth]{graphs/p_subjective_objective_v2} &	
					\includegraphics[width=0.4\linewidth]{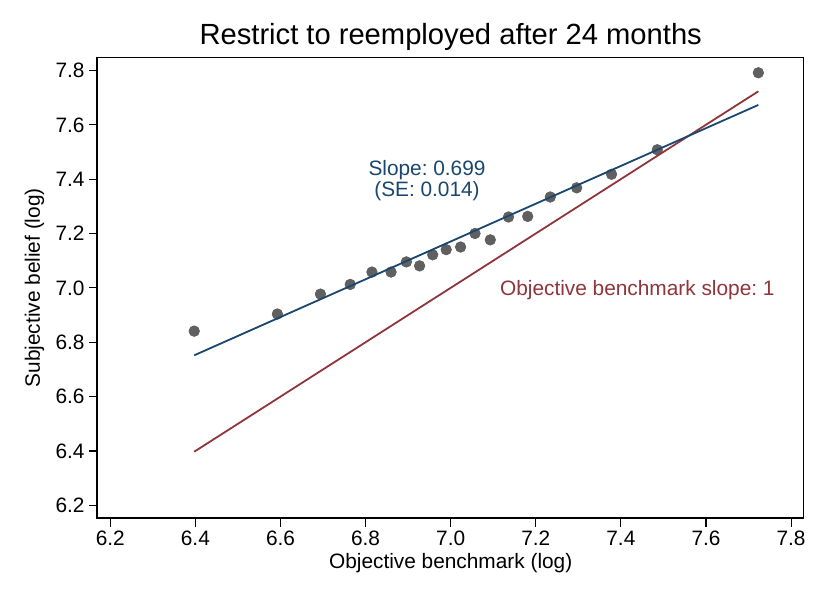} &
					
					\includegraphics[width=0.4\linewidth]{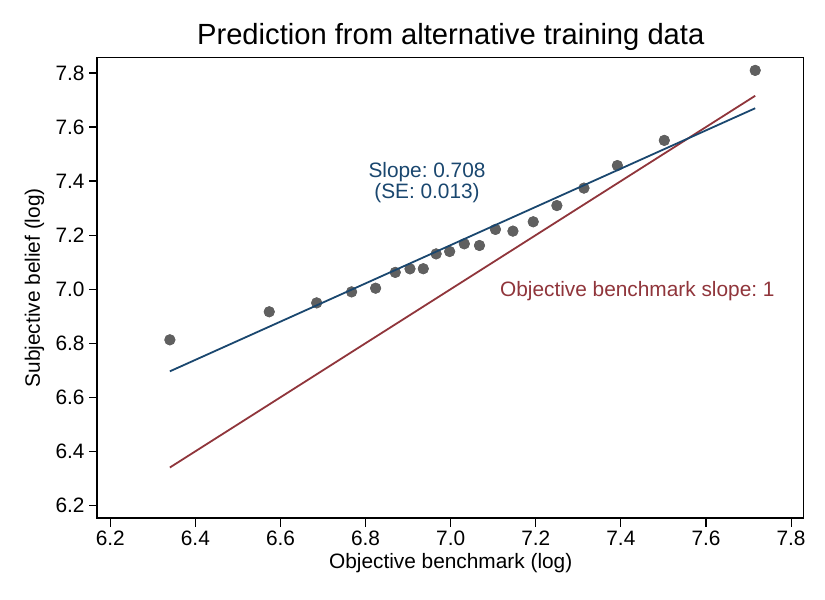} &	
					%	\textbf{D. Prediction based on reemployed after 9 months}&\textbf{E. Objective benchmark rounded to nearest 50}  & \textbf{F. Objective benchmark rounded to nearest 100}\\ 
					\includegraphics[width=0.4\linewidth]{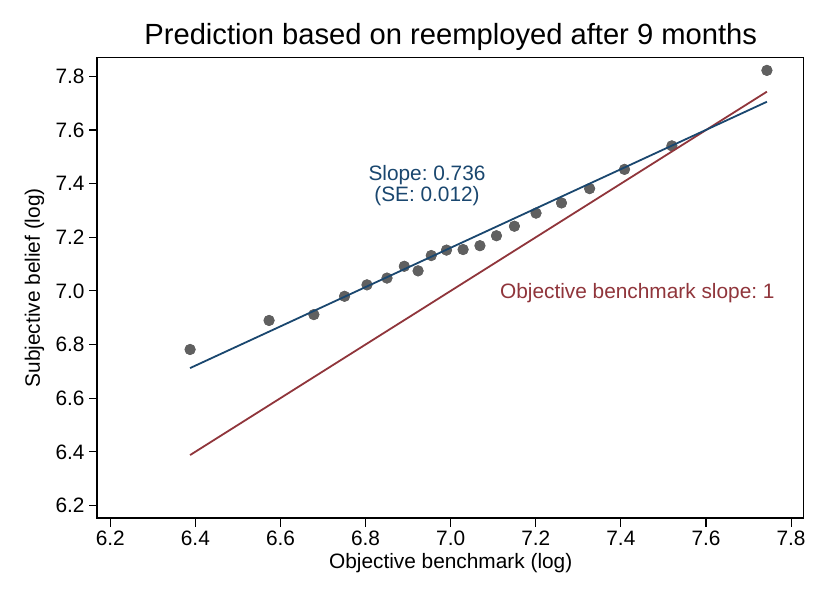} \\
					\includegraphics[width=0.4\linewidth]{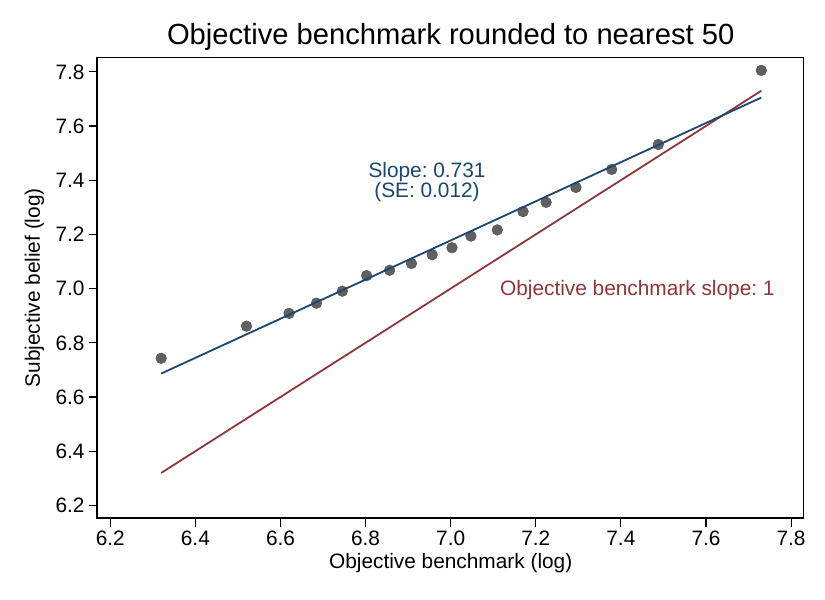} &			\includegraphics[width=0.4\linewidth]{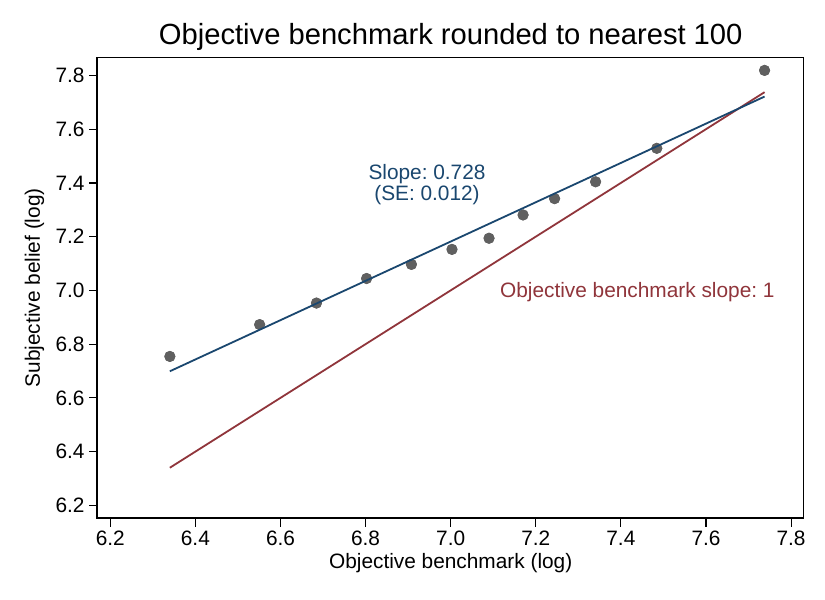} &
					%	\textbf{G. Objective benchmark rounded to nearest 250}  & \textbf{H. Restricted to searching for full-time}& \textbf{H. Restricted to searching for full-time}\\ 
					\includegraphics[width=0.4\linewidth]{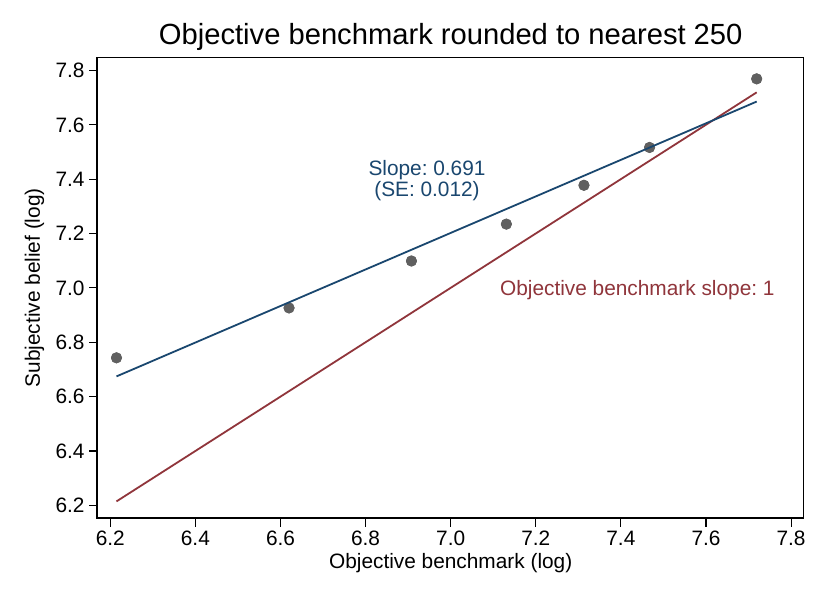} \\		\includegraphics[width=0.4\linewidth]{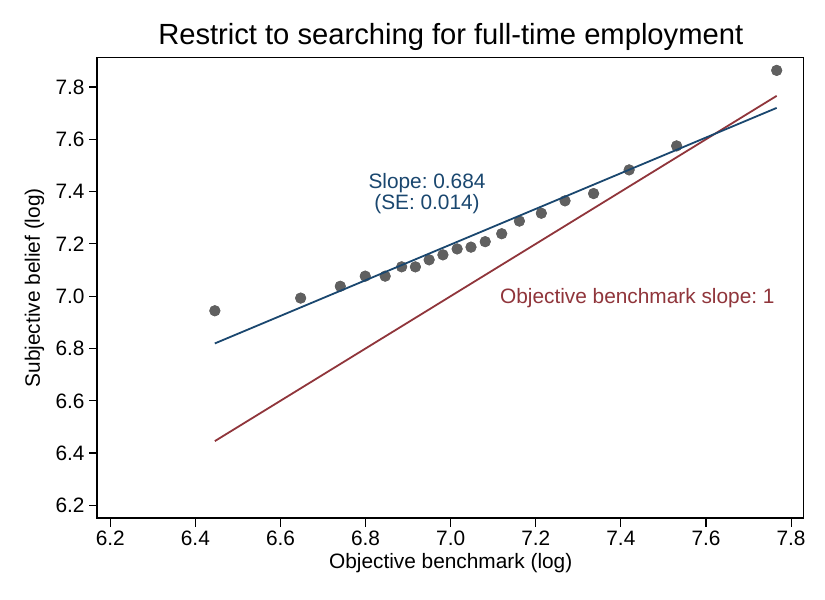} &
					\includegraphics[width=0.4\linewidth]{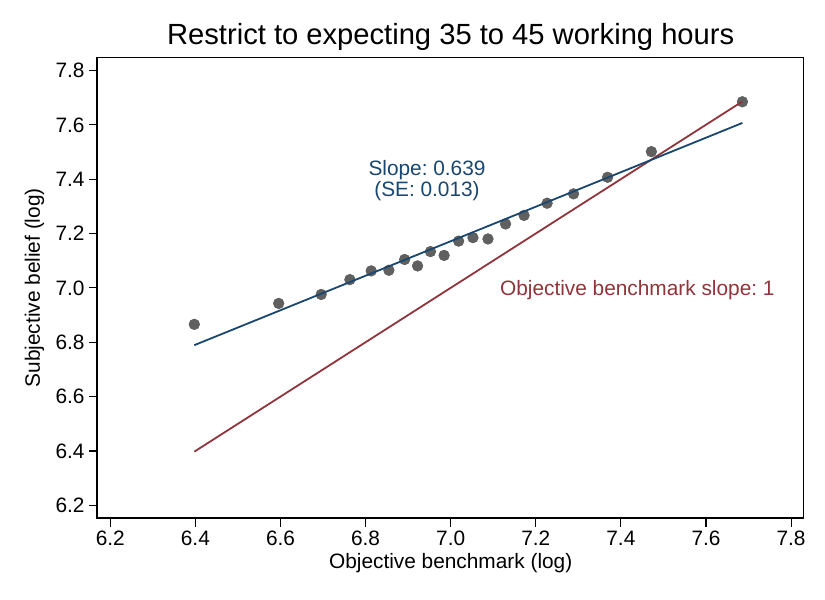}& \includegraphics[width=0.4\linewidth]{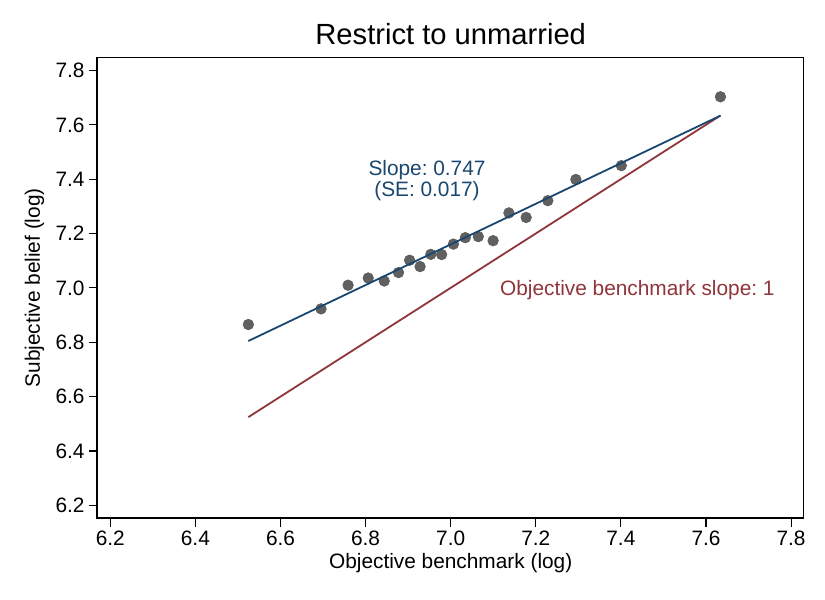}
				\end{tabular}
			\end{footnotesize}
			\begin{tablenotes} \scriptsize
				\item \textit{Note:}
				The figure depicts binned scatter plots (with 20 bins) of the robustness tests for comparisons between subjective beliefs $S_i$ and objective predictions $O_i$ for net monthly reemployment wages (both in log). The slope coefficients (with its robust standard error in parenthesis) from a regression of $S_i$ on $O_i$, correspond to the results in Table \ref{tab:subjective_objective_robust}. 
			\end{tablenotes}
	\end{threeparttable}}
\end{figure}

\vspace{2em}

%\begin{figure}[h!]
%	\caption{Asymmetric relation between subjective beliefs and objective benchmarks for wage changes \label{fig:pre_wage_asymmetric}}
%	\centerline{
%		\begin{threeparttable}
%			\begin{footnotesize}
%				\begin{tabular}{cc}
%				\includegraphics[width=0.7\linewidth]{graphs/p_subjective_objective_wagechange_survey_asymmetric}
%				\end{tabular}
%			\end{footnotesize}
%			\begin{tablenotes} \scriptsize
%				\item \textit{Note:}
%				The figure depicts a binned scatter plot (with 20 bins) for the relation between individuals' subjectively expected wage changes and objectively predicted wage changes (i.e. both in comparison to their pre-unemployment wages). Slope coefficients are estimated separately for positive and negative variation in objective wage changes. Both coefficients are significantly different from 1 (p-values  $<$ 0.001 and $=$ 0.005, respectively). $N=5,376$.
%			\end{tablenotes}
%	\end{threeparttable}}
%\end{figure}

\vspace{-2em}

\begin{figure}[h!]
	\caption{\label{fig:wave2_hist} Distribution of wage beliefs in survey waves 1 and 2}
	\centerline{
		\begin{threeparttable}
			\begin{footnotesize}
				%				\begin{tabular}{cc}
				%					\textbf{(a) Job finding within six months} & \textbf{(b) Monthly net wage in \euro}\\
				%					\includegraphics[width=0.5\linewidth]{graphs/hist_emp_wave1vs2_v2} &
				%					\includegraphics[width=0.5\linewidth]{graphs/hist_wage_wave1vs2_v2}\\
				%				\end{tabular}
				\begin{tabular}{cc}
					\includegraphics[width=0.7\linewidth]{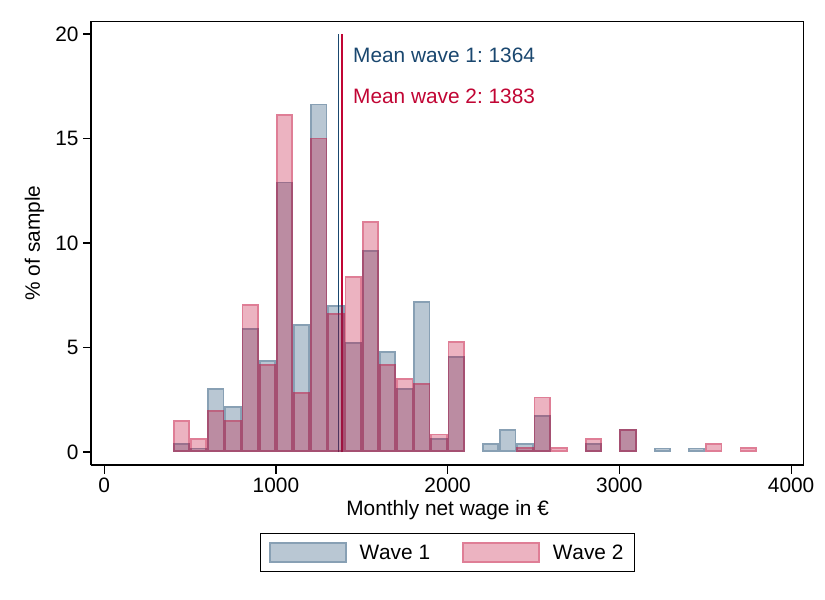}
				\end{tabular}
			\end{footnotesize}
			\begin{tablenotes} \scriptsize
				\item \textit{Note:}
				The figure shows the distribution of individuals' beliefs about net monthly reemployment wages in survey waves 1 and 2. Wave 1 was conducted 7 - 14 weeks after unemployment entry and wave 2 was collected 12 months later. The sample only includes individuals who are still in the same unemployment spell in wave 2 ($N=459$). We do not show individuals with an expected reemployment wage in wave 1 or 2 larger than 4,000\euro~($<2\%$ of sample).
			\end{tablenotes}
		\end{threeparttable}
	}
\end{figure}

\begin{comment}
\begin{figure}[h!]
\caption{\label{fig:updating_wave1} Wage belief inaccuracy in survey wave 1 across interview weeks}
\centerline{
\begin{threeparttable}
\begin{footnotesize}
\begin{tabular}{cc}
\includegraphics[width=0.7\linewidth]{graphs/p_update_wave1_bias.pdf}
\end{tabular}
\end{footnotesize}
\begin{tablenotes} \scriptsize
\item \textit{Note:} The figure shows the mean wage belief inaccuracy (defined as the log difference between subjective belief $S_i$ and objective benchmark $O_i$), together with 95\% confidence intervals, of respondents in survey wave 1 by interview week relative to unemployment entry.
\end{tablenotes}
\end{threeparttable}}
\end{figure}

\begin{figure}[h!]
	\caption{\label{fig:benefits} Comparison of UI benefits and subjective beliefs and objective benchmark}
	\centerline{
		\begin{threeparttable}
			\begin{footnotesize}
				\begin{tabular}{cc}
					\textbf{A. Subjective beliefs}  & \textbf{B. Objective benchmark}\\ 
					\includegraphics[width=0.5\linewidth]{graphs/p_beliefs_benefits} &				\includegraphics[width=0.5\linewidth]{graphs/p_obj_benefits} \\
				\end{tabular}
			\end{footnotesize}
			\begin{tablenotes} \scriptsize
				\item \textit{Note:}
				The figure compares the subjective beliefs (Panel~A) and the objective benchmark (Panel~B) and the level of unemployment benefits of the sample of survey respondents ($N = 4,647$).
			\end{tablenotes}
	\end{threeparttable}}
\end{figure}

\end{comment}

\clearpage 

\begin{table}[h!]
	\centerline{
		\begin{threeparttable}
			\caption{Summary statistics: survey respondents}\label{tab:sum_stat}
			\tabcolsep=1.3cm
			\begin{footnotesize}
				\begin{tabular}{lc}
					\hline\hline \\[-2.0ex]
					\\[-1.3ex]
					& Mean\\
					\\[-1.3ex]
					\hline
					\\[-1.3ex]
No. of observations	& 5,376 \\				
\\[-1.8ex]					
\multicolumn{2}{l}{\textbf{A. Monthly net wage in \euro}} \\ 
Subjective wage expectation ($S_i$)& 1,407\\
Objective benchmark$^{(a)}$ ($O_i$)&   1,173 \\
Realized wage$^{(b)}$ &    1,190\\
\\[-1.8ex]
Accuracy  of wage expectation: $ln(S_i) - ln(O_i)$ &      0.17\\
\\[-1.8ex]
\multicolumn{2}{l}{\textbf{B. Job finding within six months}} \\
Realized job finding rate &          0.56\\
Perceived job finding rate  \\
\quad Very likely & 0.52  \\
\quad Likely & 0.37  \\
\quad Unlikely & 0.08  \\
\quad Very unlikely & 0.03 \\
\\[-1.8ex]
\hline \hline
\end{tabular}
\end{footnotesize}
\begin{tablenotes} \scriptsize 
\item \textit{Note:} The table reports summary statistics among the sample of survey respondents.
\item $^{(a)}$Objective benchmarks are generated from realized outcomes of similar individuals observed in the administrative records as described in Section \ref{sec:data_sub}.
\item $^{(b)}$Realized wages are observed for individuals who start a regular job within 24 months after unemployment entry ($N = 4,098$).
\end{tablenotes}
\end{threeparttable}}
\end{table}

\begin{table}[htbp]\centering
	\caption{Robustness: relationship between subjective beliefs and objective benchmarks}
	\label{tab:subjective_objective_robust}
	\begin{adjustwidth}{-1.5cm}{-0cm}
		\resizebox{1.2\textwidth}{!}{
			\begin{tabular}{l*{5}{c}}
				\hline\hline \\[-2.0ex]
				\\[-2.0ex]
				& Mean & Share with    & Share with     & $\widehat{\beta}_1$ & $N$ \\
				& difference   & diff. $>$ 0.1  & diff. $<$ -0.1  &  &  \\
				\midrule
				Baseline  & 0.170 & 0.569 & 0.130    &  0.735\sym{***} (0.012) &   5,376 \\		
				[1em]		
				Restrict to individuals who are employed after 24 months &0.162 & 0.127&0.559 & 0.699\sym{***} (0.014) & 4,180 \\
				[1em]
				Prediction from alternative training data (survey period) &  0.154 & 0.536 & 0.152  &  0.708\sym{***} (0.013)  &      5,376 \\		
				[1em]
				Instrument prediction from baseline training data with   & - & - & -    &  0.740\sym{***} (0.013) &    5,376 \\	
				prediction from alternative training data 				  & & &    &   & \\	
				[1em]
				Prediction based on all individuals reentering employment  & 0.153 & 0.540 & 0.147  &  0.736\sym{***} (0.012) &  5,376 \\
				within 9 months after unemployment entry & & & \\
				[1em]
				Objective benchmark rounded to nearest 50  & 0.170 & 0.572&0.127 & 0.731\sym{***} (0.012) & 5,376 \\
				[1em]
				Objective benchmark rounded to nearest 100 & 0.170 & 0.565&0.127 & 0.728\sym{***} (0.012) & 5,376 \\
				[1em]
				Objective benchmark rounded to nearest 250 & 0.173 & 0.564& 0.143& 0.691\sym{***} (0.012) & 5,376 \\ 
				[1em]				
				Restrict to individuals searching for full-time employment &  0.179 & 0.579 & 0.110 &  0.684\sym{***} (0.014)  &    4,284 \\		
				[1em]						
				Restrict to individuals expecting between 35 and 45  &     0.165 & 0.560 & 0.118  &  0.639\sym{***} (0.013)  &  4,396 \\
				working hours \\
				[1em]
				Restrict to unmarried  &   0.155  & 0.566  & 0.117 &  0.747\sym{***} (0.017) &     3,229 \\
				[0.5em]			
				\hline\hline
				\multicolumn{6}{l}{
					\begin{minipage}{1.2\linewidth} \footnotesize \smallskip 
						\textit{Note:} Robustness tests for comparisons between subjective beliefs $S_i$ and objective predictions $O_i$ for net monthly reemployment wages (both in log). The table reports the mean difference between $S_i$ and $O_i$, the shares of individuals who overestimate and underestimate wages by more than 10\% (difference $>$ 0.1 / $<$ 0.1), the slope coefficient $\widehat{\beta}_1$ (with its robust standard error in parenthesis) from a regression of $S_i$ on $O_i$, and the number of individuals in the sample. ***/**/* indicate statistical significance at the 1\%/5\%/10\%-level for a test of the null hypothesis that $\beta_1 = 1$.
					\end{minipage}
				}\\	
			\end{tabular}
		}
	\end{adjustwidth}
\end{table}

\begin{table}[h!]
	\centerline{
		\begin{threeparttable}
			\caption{Descriptive evidence on labor market implications} \label{tab:bias}
			\tabcolsep=0.8cm
			\begin{footnotesize}
				\begin{tabular}{lcccc}
					\hline\hline 
						\\[-1.3ex]
		Dependent variable & \multicolumn{2}{c}{Log no. of job} & \multicolumn{2}{c}{Log realized net} \\
& \multicolumn{2}{c}{applications$^{(a)}$} & \multicolumn{2}{c}{monthly wage$^{(b)}$} \\ 
	\\[-2.0ex]
\cmidrule(lr){2-3} \cmidrule(lr){4-5}
	\\[-2.0ex]
		& (1) & (2) 	& (3) & (4) \\
		\\[-2.0ex]
		\midrule
		\\[-2.0ex]		
\mc{5}{l}{Accuracy of wage expectations: $log(S_i) - log(O_i)$}\\
\quad linear&        0.163\sym{***}&       0.249\sym{***}&   0.306\sym{***}&       0.364\sym{***}\\
				   &        \raisebox{.7ex}[0pt]{\scriptsize(0.057)}         &      \raisebox{.7ex}[0pt]{\scriptsize(0.075)}       &      \raisebox{.7ex}[0pt]{\scriptsize(0.022)}         &      \raisebox{.7ex}[0pt]{\scriptsize(0.030)}    \\
\quad squared&             &      -0.171  &                  &      -0.129\sym{***} \\
							&            &     \raisebox{.7ex}[0pt]{\scriptsize(0.107)}  &                    &     \raisebox{.7ex}[0pt]{\scriptsize(0.047)}         \\
	\\[-2.0ex]
No. of observations          &          5,376      &    5,376       &          4,098      &   4,098             \\
$R^2$           &       0.023         &       0.024     &       0.490         &       0.491      \\
\\[-1.3ex]
\midrule 
\\[-1.3ex]
		Dependent variable & \multicolumn{2}{c}{Perceived job finding} & \multicolumn{2}{c}{Realized job finding} \\
		& \multicolumn{2}{c}{rate (1=very high)$^{(c)}$} & \multicolumn{2}{c}{rate (1=yes)$^{(d)}$} \\ 
	\\[-2.0ex]
\cmidrule(lr){2-3} \cmidrule(lr){4-5}
\\[-2.0ex]
	& (5) & (6) 	& (7) & (8) \\
	\\[-2.0ex]
\midrule
\\[-2.0ex]
\mc{5}{l}{Accuracy of wage expectations: $log(S_i) - log(O_i)$}\\
\quad linear&       0.166\sym{***}&       0.158\sym{***} &       -0.046\sym{*}  &       0.036\\
				&        \raisebox{.7ex}[0pt]{\scriptsize(0.027)}         &     \raisebox{.7ex}[0pt]{\scriptsize(0.038)}    &      \raisebox{.7ex}[0pt]{\scriptsize(0.027)}         &     \raisebox{.7ex}[0pt]{\scriptsize(0.037)}         \\
\quad squared&             &       0.015 &      &      -0.163\sym{***}    \\
							&            &     \raisebox{.7ex}[0pt]{\scriptsize(0.055)} &      &     \raisebox{.7ex}[0pt]{\scriptsize(0.053)}               \\
	\\[-2.0ex]
No. of observations        &          4,953      &   4,953    &          5,376      &    5,376                  \\
$R^2$           &       0.066         &       0.066        &       0.028         &       0.029   \\
	\\[-1.3ex]
\hline\hline
\end{tabular}
	\end{footnotesize}
\begin{tablenotes} \scriptsize 
\item \textit{Note:} The table reports the results of OLS regressions. The explanatory variable refers to the accuracy of job seekers' wage expectations defined as the log difference between the subjective belief $S_i$ and the objective benchmark $O_i$. In all specifications, we control for socio-demographic characteristics and objective benchmarks. Robust standard errors are shown in parentheses. ***/**/* indicate statistical significance at the 1\%/5\%/10\%-level.
\item $^{(a)}$ The dependent variable is the log number of job applications sent since unemployment entry.
\item $^{(b)}$ The dependent variable is the first monthly net wage of individuals who start a regular job within 24 months after entry into unemployment.
\item $^{(c)}$ The dependent variable is an indicator of whether the individual reports the perceived six-month-ahead job finding rate to be `very high'.
\item $^{(d)}$ The dependent variable is an indicator of whether the individual actually starts a regular job within the next six months.\end{tablenotes}
\end{threeparttable}}
\end{table}

\clearpage
\newpage
\renewcommand{\thesection}{B}
\renewcommand{\thetable}{B.\arabic{table}}
\setcounter{table}{0}
\renewcommand{\thefigure}{B.\arabic{figure}}
\setcounter{figure}{0}

\section{Details on Objective Benchmarks} \label{app:data}	

%\subsection{Generating Objective Benchmarks} \label{app:lasso}	

In the following, we present details about the prediction of objective benchmarks for job seekers' reemployment wages. 

\paragraph{Covariates:} Monthly wages are modeled as a function of covariates which are pre-determined at the time of unemployment entry. We exploit a rich set of characteristics that are available in both the \textit{IZA/IAB Linked Evaluation Dataset} (LED) and the \textit{IZA/IAB Administrative Evaluation Dataset} (AED). These include sociodemographic information (e.g. gender, age, education, family status), characteristics of the last job before unemployment (wage, part- versus full-time job), detailed information on the labor market biography in the last ten years (e.g. months employed, unemployed, and in labor market programs), and local labor market characteristics (unemployment rate, East versus West German residency). To allow for a flexible functional form, we also use third-order polynomial terms of all continuous variables and first-order interaction terms of all variables with some important characteristics (gender, age, education, East German residency, last wage). See Table \ref{tab:list_covariates} for a complete list of all the 717 included covariates. 

\paragraph{LASSO regression:} To address the high-dimensional nature of the data and to avoid over-fitting, we estimate LASSO regressions, which add a regularization term to the objective function and shrink some coefficients to zero. Specifically, we estimate a linear LASSO regression for wages and a logistic LASSO regression for job finding probabilities.\footnote{We use the Stata commands \texttt{lasso2} from the \texttt{lassopack} developed by \cite{Ahrens2018, Ahrens2020}.} We optimized the regularization parameter $\lambda$ using five-fold cross-validation in the training data and considering the following values for $\lambda$: 0.001, 0.01, 0.1, 1, 10, 100, 1000. In the estimated models, 66\% of all covariates are selected.

\paragraph{Prediction quality:} In Table~\ref{tab:performance}, we evaluate the out-of-sample predictive performance of the LASSO model. Results are presented for the different training samples (AED 01/2005 - 06/2007 or 80\% of AED 06/2007 - 06/2008) and test samples (80\% of AED 06/2007 - 06/2008 or LED). We calculate the out-of-sample $R^2$ obtained from a regression of realized gross wages on predicted wages in the test data. The predictions explain between 48\% and 53\% of the out-of-sample variation. Note that the predictive performance is similar for the two different training samples. The correlation of predictions between both samples is also very high (0.976). 

\begin{figure}[h!]
	\caption{Relation between realized wages and objective benchmarks \label{fig:obj_vs_realized}}
	%\vspace{-0.5cm}
	\centerline{
		\begin{threeparttable}
			\begin{footnotesize}
				\begin{tabular}{cc}
				\includegraphics[width=0.7\linewidth]{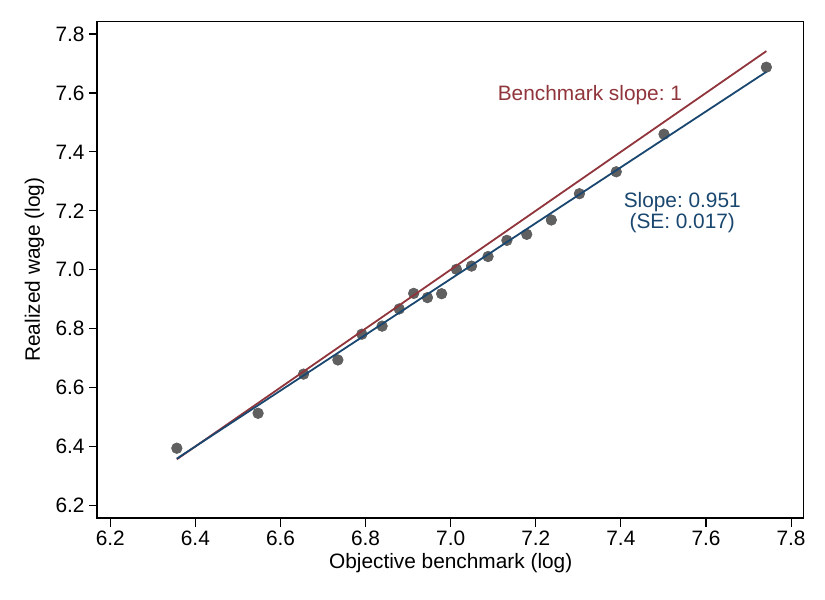}
				\end{tabular} 
			\end{footnotesize}
			\begin{tablenotes} \scriptsize
				\item \textit{Note:}
				The figure depicts a binned scatter plot (with 20 bins) for the relation between individuals' realized reemployment wages and objective benchmarks for reemployment wages. $N=4,098$.
				%The slope is significantly different from 1 (p-values  $<$ 0.001).
			\end{tablenotes}
	\end{threeparttable}}
\end{figure}

Results in Table~\ref{tab:performance} are based on gross wages. For our analyses, we convert the predicted gross wage into net terms (see procedure below). For the converted net wage predictions in the LED sample, we can compare the explanatory power of objective predictions with that of respondents' subjective beliefs. The results in Table~\ref{tab:performance_objVSsub} demonstrate that our lasso model is better at predicting realized wages than job seekers themselves are: while job seekers' beliefs explain 34\% of the variation in realized net wages, the objective benchmarks explain 45\%.\footnote{This result is in line with the evidence provided by \cite{Berg2022} who find that machine learning predictions have higher explanatory power for reemployment probabilities than job seekers' self assessments.} Figure~\ref{fig:obj_vs_realized} also examines the fit over the distribution of our predictions. The objective benchmarks predict realized wages in the survey sample with a slope coefficient very close to one, indicating that there are not systematic prediction errors over the objective benchmark distribution.

In Table~\ref{tab:realized_personality}, we also investigate whether job seeker characteristics that are unobserved in the administrative data are likely to affect the validity of our objective benchmarks. For that we regress log realized wages on a set of personality traits, including locus of control, conscientiousness, openness, extraversion, and neuroticism, which are observed in the survey sample. The personality traits show strong correlations with realized wages. However, when we condition on the objective benchmark measure, the correlations vanish almost entirely. An F-test fails to reject joint significance of all personality traits and, individually, only locus of control remains significant. These results indicate the saturation of our prediction model based on the rich set of covariates in the administrative data.

\paragraph{Converting gross to net wages:} Wage beliefs in the survey are elicited in net terms, while the administrative data provide realized wages in gross terms. We therefore convert gross into net wages by deducting social security contributions and wage taxes.

Wage taxes are withheld by the employer and deducted from the monthly wage payment. They qualify as a pre-payment of the income tax in case the employee files an annual income tax declaration. We do not perform a complete income tax calculation since we do not have information on individual-specific deductions and other income sources. Moreover, individuals most likely think about their monthly payroll when asked about their net wage. Rather, we calculate withheld wage taxes utilizing contribution and tax schedules of 2008 and taking into account variation in rates according to partnership status, number of children, age, and East versus West German residency. 

Exemption thresholds in the wage tax schedule depend on the tax class of the individual. While single individuals without children are always in class I and single parents are always in class II, married couples may choose between a combination of classes IV/IV or III/V. With IV/IV, the standard exemption threshold is applied to both spouses, whereas with III/V the higher-earning spouse obtains twice the standard exemption rate while the second earner is already taxed at lower earnings. Although we do not directly observe the chosen tax classes for married couples, we can infer them from the observed UI benefit payments. UI benefits are generally calculated by the following formula: Monthly benefit = 0.6 $\times$ (Average monthly gross wage in last 12 months before unemployment - Wage tax - SolZ - Social security payment). Thus, the relation between previous gross wages and UI benefits depends on the chosen tax class. We exploit this relation by calculating the hypothetical benefits under tax classes III, IV, and V and then choose the tax class that minimizes the difference between the actual observed and the hypothetical benefit payments. The derived tax classes match well-known descriptives about taxation of married couples in Germany. For instance, tax class V is chosen more frequently by women than men and this type of splitting is more prevalent in West than in East Germany.

\begin{table}[htbp]\centering
	\centerline{
	\begin{threeparttable}
		\caption{Summary statistics: LED versus AED samples}\label{tab:sumstat_led_aed}
		\begin{footnotesize}
			\begin{tabular}{lccc}
				\hline\hline \\[-2.0ex]
				\\[-2.0ex]
				& LED & AED & AED  \\ 
				& \hspace{8em} & 01/2005 - 05/2007  & 06/2007 - 05/2008  \\ 			
				\midrule
				No. of observations & 5,376 & 84,617 & 21,715 \\[1em]
				%\multicolumn{4}{l}{\textbf{Reemployment outcomes}}\\
				Reemployment wage (gross, per month)$^{(a)}$ 	& 1,782&     1,704&      1,716   \\
				%Reemployed in next six months 					&  0.56&        0.46 &           0.50        \\
				[1em]
				\multicolumn{4}{l}{\textbf{Socio-demographic characteristics}}\\ 
				Female  								&        0.41&        	 0.35&             0.37   \\
				Age   									&       36.16&      	35.11&            34.90 \\
School leaving degree\\
\quad lower secondary degree  			&        0.31&         	 0.38&            0.35 \\
\quad middle secondary degree  			&        0.44&         	 0.39&            0.40\\
\quad upper secondary degree					&        0.23&     		 0.18&          0.21 \\
Further education\\
\quad vocational certificate &      0.72&           0.68&             0.68    \\
\quad university degree			&        0.20&           0.16&         0.19 \\
				German citizen 							&        0.94&      	 0.91&         0.91 \\
				Married    								&        0.39&           0.45&             0.43 \\
Number of children\\
\quad one child  					&        0.18&           0.16&            0.17 \\
\quad two or more children  				&        0.12&           0.14&          0.13\\	
				[1em]
				\multicolumn{4}{l}{\textbf{Last job}}\\ 	
				Wage in last job (gross, per month) 	&    	1,726&      1,688&       1,696\\
				Last job was quit by individual 		&        0.06&         0.07&         0.09\\
				[1em]
				\multicolumn{4}{l}{\textbf{Labor market history}}\\ 		
				\# of employers in last 2 years &        2.55&          1.49&         1.61 \\			
				\# of employers in last 10 years &       4.64&          3.13&        3.45 \\	
				\# of UE spells in last 2 years &        1.41&          0.59&         0.44 \\
				\# of ALMP programs in last 2 years &        0.42&      0.17&         0.21 \\		
				Duration of last UE spell in months &        5.09&      5.34&         5.80 \\		
				\# of months employed in year t-1 &        8.64&       10.74&        10.92 \\
				\# of months employed in year t-2 &        7.86&        9.27&         9.14\\		
				\# of months employed in year t-3 &        7.50&        8.38&         7.73 \\
				\# of months employed in years t-4 to t-10    &       46.12&       43.18&    40.20 \\					
				\# of months unemployed in year t-1 &        0.79&      0.86&         0.61 \\
				\# of months unemployed in year t-2 &        1.07&      1.57&        1.49 \\		
				\# of months unemployed in year t-3 &        1.29&      1.75&       2.19 \\
				\# of months unemployed in years t-4 to t-10  &       8.81&      8.37&      10.42\\	
				\# of months in ALMP in year t-1 &        0.60&     0.22&         0.25 \\
				\# of months in ALMP in year t-2 &        0.68&     0.53&        0.57 \\		
				\# of months in ALMP in year t-3 &        0.63&     0.61&          0.61 \\
				\# of months in ALMP in years t-4 to t-10 &        1.35&       1.20&        1.42 \\	
				Average wage in year t-1 (quintiles) &        3.17&    3.25&      3.19 \\
				Average wage in year t-2 (quintiles) &        2.97&    3.00&      2.89 \\		
				Average wage in year t-3 (quintiles) &        2.78&    2.81&      2.59 \\			
				Average wage in years t-4 to t-10 (quintiles) &       17.88&      17.01&    16.41 \\	
				[1em]
				\multicolumn{4}{l}{\textbf{Local labor market}}\\ 	
				West, UE rate $<$3\%      	&        0.02&       0.00&    0.01 \\	
				West, UE rate 3-6\%    		&        0.29&       0.07&    0.26   \\
				West, UE rate 6-9\%       	&        0.25&       0.24&    0.26 \\
				West, UE rate $>$9\%    	&        0.11&       0.39&    0.18 \\
				East, UE rate $<$12\%     	&        0.07&       0.00&    0.04 \\
				East, UE rate 12-14\%    	&        0.10&       0.01&    0.08\\
				East, UE rate 14-16\%    	&        0.10&       0.05&    0.12\\
				East, UE rate $>$16\%     	&        0.04&       0.23&    0.05\\	 
				\hline\hline
			\end{tabular}
\end{footnotesize}
\begin{tablenotes} \scriptsize 
\item \textit{Note:} The table compares the mean characteristics of the survey sample (LED) with those of the two adminstrative samples used for the prediction of objective benchmarks (AED). 
\item $^{(a)}$Realized wages are observed for individuals who start a regular job within 24 months after unemployment entry.
\end{tablenotes}
\end{threeparttable}}
\end{table}

\clearpage
\begin{table}[p]
	\centerline{
		\begin{threeparttable}
			\vspace{-25pt}
			\caption{List of covariates used to generate objective benchmarks}\label{tab:list_covariates}
			\tabcolsep=0.25cm
			\begin{footnotesize}
				\begin{tabular}{ll}
					\hline\hline 
		\\[-2.0ex]
		Type & Covariate \\
		\\[-1.3ex]
		\multicolumn{2}{l}{\textbf{Socio-demographic characteristics}}\\ 
			Continuous 	& Age \\
			Categorical & Female \\
			& School degree (4 categories: none, lower sec. degree, middle sec. degree, upper sec. degree) \\
			& Further education (3 categories: none, vocational certificate, university degree) \\
			& German citizen \\
			& Married \\
			& \# of children (3 categories: 0, 1, $\geq$ 2) \\
			[1em]
			\multicolumn{2}{l}{\textbf{Last job}}\\ 
			Continuous 	& Wage in last job (gross, per month) \\		
			Categorical & Last job was quit by individual \\	
			[1em]
			\multicolumn{2}{l}{\textbf{Labor market history}}\\ 				
			Continuous 	& Duration of last UE spell in months \\		
			& \# of months employed in year t-1 \\
			& \# of months employed in year t-2 \\		
			& \# of months employed in year t-3 \\
			& \# of months employed in years t-4 to t-10 \\					
			& \# of months unemployed in year t-1 \\
			& \# of months unemployed in year t-2 \\		
			& \# of months unemployed in year t-3 \\
			& \# of months unemployed in years t-4 to t-10 \\	
			& \# of months in ALMP in year t-1 \\
			& \# of months in ALMP in year t-2 \\		
			& \# of months in ALMP in year t-3 \\
			& \# of months in ALMP in years t-4 to t-10 \\										
			Categorical & \# of employers in last 2 years (5 categories) \\			
			& \# of employers in last 10 years (5 categories) \\	
			& \# of UE spells in last 2 years (5 categories) \\
			& \# of ALMP programs in last 2 years (5 categories) \\
			& Average wage in year t-1 (5 categories) \\
			& Average wage in year t-2 (5 categories) \\		
			& Average wage in year t-3 (5 categories) \\			
			& Average wage in years t-4 to t-10 (5 categories) \\					
			& All months regularly employed in year t-1 \\
			& Zero months regularly employed in year t-2 \\		
			& All months regularly employed in year t-2 \\		
			& Zero months regularly employed in year t-3 \\
			& All months regularly employed in year t-3 \\
			& Zero months regularly employed in years t-4 to t-10 \\		
			& All months regularly employed in years t-4 to t-10 \\							
			& Zero months unemployed in year t-1 \\
			& Zero months unemployed in year t-2 \\		
			& Zero months unemployed in year t-3 \\
			& Zero months unemployed in years t-4 to t-10 \\		
			& Zero months in ALMP in year t-1 \\
			& Zero months in ALMP in year t-2 \\		
			& Zero months in ALMP in year t-3 \\
			& Zero months in ALMP in years t-4 to t-10 \\
			[1em]
			\multicolumn{2}{l}{\textbf{Local labor market}}\\ 
			Categorical & Unemployment rate in employment agency district at time of UE entry \\	
			& (8 categories: West $<$3\%, 3-6\%, 6-9\%, $>$9\%; East $<$12\%, 12-14\%, 14-16\%, $>$16\%)		\\
			[1em]
			\multicolumn{2}{l}{\textbf{Timing of unemployment spell}}\\ 	
			Categorical & Calendar month of entry into unemployment (12 categories) \\
	\\[-2.0ex]
\hline \hline
\end{tabular}
\end{footnotesize}
\begin{tablenotes} \scriptsize
\item \textit{Note:} The table reports the list of covariates used in the lasso regression to predict reemployment wages. We also include second- and third-order polynomials of all continuous variables, as well as interaction terms of all variables with \textit{female}, \textit{age}, \textit{upper secondary degree}, \textit{East German residency}, and the \textit{monthly gross wage in the last job}, yielding a total of 717 covariates included.
\end{tablenotes}
\end{threeparttable}}
\end{table}

\clearpage
\begin{table}[h!]
	\centerline{
		\begin{threeparttable}
	\caption{Prediction performance based on out-of-sample $R^2$}
	\label{tab:performance}
\tabcolsep=0.5cm
	\begin{footnotesize}
	\begin{tabular}{lcccc}
			\hline\hline \\[-2.0ex]
		 & \mc{4}{c}{Test dataset$^{(a)}$}\\
		 \\[-2.0ex]
		 \cline{2-5}
		 \\[-1.3ex]
		 & \multicolumn{2}{c}{Administrative data} & \multicolumn{2}{c}{Survey data} \\
		  \cmidrule(lr){2-3} \cmidrule(lr){4-5}
		Training dataset$^{(b)}$ & Baseline & Robustness & Baseline & Robustness \\
		\midrule \\[-2.0ex]
		%&\multicolumn{4}{c}{\textbf{Job finding}} \\ \cmidrule(lr){2-5}
		%ROC-AUC  & .657	& .653 & .632 & .638 \\
		%[1em]
		%&\multicolumn{4}{c}{\textbf{Wage}} \\ \cmidrule(lr){2-5}
		Out-of-sample $R^2$ 	   	& 0.525	& 0.516 	& 0.511 	& 0.481 \\
		%		Net monthly wage		&	    & 	   	& .491 	& .469 \\
		%		Net wage (log)	& 		& 		& .384  & .349 \\
	\\[-2.0ex]
	\hline \hline
\end{tabular}
\end{footnotesize}
\begin{tablenotes} \scriptsize 
\item \textit{Note:} The table reports the out-of-sample $R^2$, which is obtained based on a regression of realized reemployment wages as observed in the respective test dataset on predictions of objective benchmarks generated in the training dataset.
\item $^{(a)}$The administrative test dataset includes a 20\% sample of all entries into unemployment between June 2007 and May 2008, while the survey test dataset includes all individuals observed in the matched survey-administrative data.
\item $^{(b)}$The baseline training dataset includes all entries into unemployment between January 2005 and May 2007 as observed in the administrative records. As a robustness check, we use an 80\% sample of all entries into unemployment between June 2007 and May 2008 (i.e. the same period during which the survey was conducted) excluding observations from the respective test dataset.
\end{tablenotes}
\end{threeparttable}}
\end{table}

\vspace{3em}
\begin{table}[h!]
	\centerline{
		\begin{threeparttable}
			\caption{Predictive power of objective benchmarks and subjective beliefs}
		\label{tab:performance_objVSsub}
			\tabcolsep=1.1cm
			\begin{footnotesize}
				\begin{tabular}{lcc}
					\hline\hline 
					\\[-2.0ex]
Dependent variable & \mc{2}{c}{Log realized net monthly wage} \\
\\[-2.0ex]
		& (1) & (2) \\
\\[-2.0ex]
\hline
\\[-2.0ex]
		Objective benchmark (log)&       0.951\sym{***}&                     \\
		&    \raisebox{.7ex}[0pt]{\scriptsize(0.017)}         &             \\
		[1em]
Subjective belief (log)     &    &   0.758\sym{***}\\
		&                     &    \raisebox{.7ex}[0pt]{\scriptsize(0.018)}       \\
%		\midrule
		No. of observations           &    4,098         &        4,098                \\
		$R^2$          &   0.454         &       0.335                \\
		%		ROC-AUC		  & 0.632	& 0.606	& - & - \\
	\\[-2.0ex]
\hline \hline
\end{tabular}
\end{footnotesize}
\begin{tablenotes} \scriptsize 
\item \textit{Note:} The table reports the results of an OLS regressions of log realized reemployment wages on objective benchmarks and subjective wage expectations, respectively. The sample includes survey respondents as observed in the matched survey-administrative data who found a new job within 24 months after the beginning of the unemployment spell. Standard errors reported in parenthesis. ***/**/* indicate statistical significance at the 1\%/5\%/10\%-level. 
\end{tablenotes}
\end{threeparttable}}
\end{table}

\begin{table}[p]
	\centerline{
		\begin{threeparttable}
			\vspace{-55pt}
			\caption{Personality traits and realized wages}\label{tab:realized_personality}
			\tabcolsep=0.25cm
			\begin{footnotesize}
				\begin{tabular}{lcc}
					\hline\hline \\[-2.0ex]
					%\\[-2.0ex]					
					& \mc{2}{c}{Log realized net monthly wage} \\
					 & (1) & (2)\\ \hline \\[-2.0ex]
%\mc{2}{l}{\textbf{Personality traits}}\\
%[1em]
Internal locus of control                        &       0.039\sym{***}&       0.014\sym{**}\\
                                  &     (0.007)         &     (0.006)      \\
[1em]
Conscientiousness                     &      -0.001         &       0.002         \\
                                &     (0.007)         &     (0.005)         \\
[1em]
Openness                      &       0.029\sym{***}&       0.004         \\
                                 &     (0.007)                &     (0.005)         \\
[1em]
Extraversion                      &      -0.029\sym{***}&      -0.004          \\
                                 &     (0.008)         &     (0.006)      \\
[1em]
Neuroticism                     &      -0.041\sym{***}&       0.002           \\
                                 &     (0.007)              &     (0.005)         \\
[1em]
Objective benchmark $O_i$                       &                     &       0.941\sym{***}\\
            &                        &     (0.017)         \\
\hline
\\[-2.0ex]
No. of observations                &    3,959             &    3,959        \\
$R^2$          &             0.024         &             0.453         \\
%F-Test &17.41 & 1.33 \\
p-value joint significance & $<$0.001 & 0.247 \\
\hline\hline
\end{tabular}
\end{footnotesize}
\begin{tablenotes} \scriptsize
\item \textit{Note:} The table reports the results of OLS regressions of log realized net monthly reemployment wages on personality traits and objective benchmarks for reemployment wages. It also reports p-values for the joint significance of all personality traits. Robust standard errors are shown in parentheses. ***/**/* indicate statistical significance at the 1\%/5\%/10\%-level.
\end{tablenotes}
\end{threeparttable}}
\end{table}

\vspace{2em}

\renewcommand{\thesection}{C}
\renewcommand{\thetable}{C.\arabic{table}}
\setcounter{table}{0}
\renewcommand{\thefigure}{C.\arabic{figure}}
\setcounter{figure}{0}

\section{Details on Sanction Analysis} \label{app:sanc}
		
\subsection{Econometric strategy}

\begin{figure}[h!]
	\caption{\label{fig:sanction_intensity} Distribution of local sanction intensity across survey respondents}
	\centerline{
		\begin{threeparttable}
			\begin{footnotesize}
				\begin{tabular}{cc}
					\includegraphics[width=0.7\linewidth]{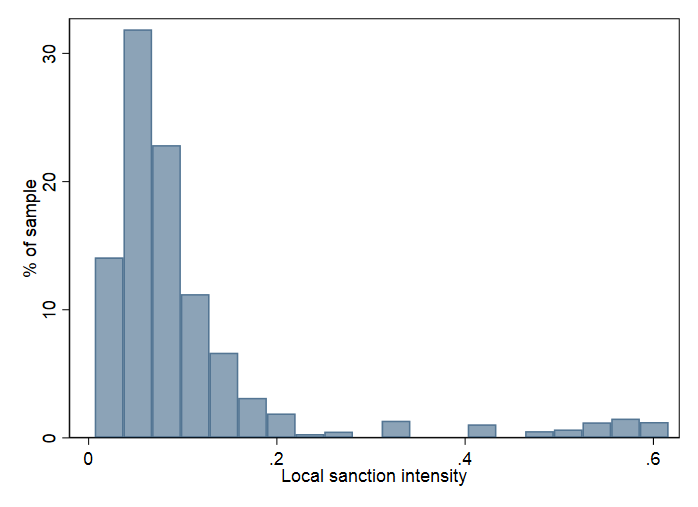}
				\end{tabular}
			\end{footnotesize}
			\begin{tablenotes} \scriptsize
				\item \textit{Note:} The figure shows the distribution of local sanction intensity across all survey respondents ($N = 5,669$).
			\end{tablenotes}
	\end{threeparttable}}
\end{figure}

\begin{figure}[h!]
	\caption{\label{fig:map} Distribution of local sanction intensity across LEA districts}
	\centerline{
		\begin{threeparttable}
			\begin{footnotesize}
				\begin{tabular}{cc}
					\includegraphics[width=0.7\linewidth]{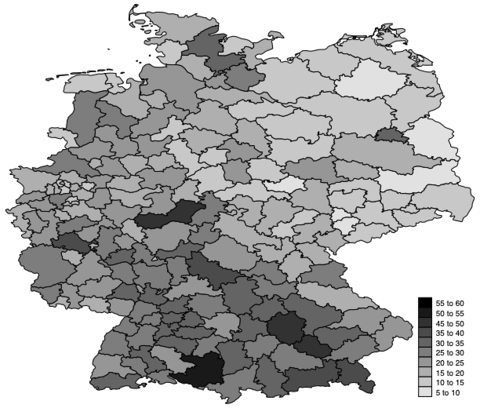}
				\end{tabular}
			\end{footnotesize}
			\begin{tablenotes} \scriptsize
				\item \textit{Note:} The figure shows the geographical distribution of local sanction intensity across LEA districts in Germany.
			\end{tablenotes}
	\end{threeparttable}}
\end{figure}

To capture variations in the local sanction regime, we utilize regional data on the annual number of benefit sanctions imposed in each of the 178 LEA districts (indexed by $j$) and normalize this information by the average annual stock of unemployed workers in each LEA district. The resulting sanction intensity, $SI_j$, can be linked to the administrative and survey data, as explained in Section~\ref{sec:data}, both of which include identifiers for job seekers' place of residence.\footnote{Due to data security restrictions, we are unable to utilize regional identifiers for the linked survey-administrative data in our analysis. Consequently, in this section, we rely on the survey and administrative data without linking them at the individual level. This requires us to re-estimate the objective benchmarks using a reduced set of covariates available in both the survey and administrative records. This includes socio-demographic characteristics, previous wage, regional information, and month of entry into unemployment. Despite this adjustment, our model demonstrates strong out-of-sample predictive power (with an $R^2$ of 0.39), and the re-estimated objective predictions closely align with the measure employed in the previous sections ($\rho=0.75$). \label{fn:region}} To ensure that our estimation sample does not contribute to the sanction intensity measure, we rely on the corresponding numbers as observed in the year before a job seeker entered unemployment. In the Appendix, we illustrate the distribution of the sanction intensity across survey respondents (see Figure~\ref{fig:sanction_intensity}), as well as LEA districts in Germany (see Figure~\ref{fig:map}).

%\footnote{However, it should be noted that we cannot use the linked version of both datasets because the latter does not include a regional identifier for privacy reasons. Therefore, our analysis relies on two samples of individuals who entered unemployment between June 2007 and May 2008, including 5,669 survey respondents and 25,654 individuals observed in the administrative data. Besides that, we employ the same sample restrictions as in our previous analyses. In order to obtain an objective benchmark for survey respondents' wage potentials, we estimate a prediction model that relies on a reduced set of covariates observed in both the survey and administrative data (sociodemographics, last wage, regional unemployment rate, month of unemployment entry). The out-of-sample predictive power remains strong ($R^2$ of XX) and in the linked survey-administrative data we observe a correlation coefficient of XX between our baseline objective benchmark with the predictions from the reduced set of covariates. This makes us confident that we still exploit a meaningful objective benchmarks for job seekers in the (unlinked) survey sample.} 
%The resulting measure is winsorized at the top and the bottom percentile.

While the local sanction intensity serves as a proxy for the personal risk of being exposed to a benefit sanction, LEA districts imposing more sanctions might face a different composition of the unemployed workforce. This makes it unlikely that a simple regression of job seekers' outcomes on the local sanction intensity will identify the causal effect of job seekers' personal sanction risk. Therefore, we exploit discontinuities with respect to the sanction intensity along the administrative borders of the LEA districts \citep[similar to][]{dube2010minimum,Caliendo2022}. Specifically, we estimate border-pair fixed-effects models of the following form:
\begin{align}
	Y_{ijb} &= \alpha + \delta SI_{j} + \beta X_{i} + \phi R_{j} + \kappa_b + \varepsilon_{ijb}, \label{est:base}
\end{align}
where $i$ denotes the individual job seeker, $j$ the LEA district in which the individual is located at the beginning of the unemployment spell, and $b$ a pair of bordering LEA districts such that $\kappa_b$ denotes the border-pair fixed effects for any combination of two neighboring LEA districts. Since one LEA district usually has several neighboring districts, an individual living in region $j$ can belong to different sets of boarder pairs $b$ and therefore enters the estimation multiple times (depending on the number of neighboring regions). Therefore, we use sampling weights referring to the inverse of the number of neighboring LEA districts. The parameter of interest $\delta$ identifies the effect of sanction intensity on the outcome variables $Y$ by comparing individuals living in similar, neighboring LEA districts but facing varying risks of being sanctioned. Moreover, $R_j$ captures regional characteristics including the local unemployment rate, vacancy rate, gross domestic product, industry structure, and federal state fixed effects, and $X_i$ accounts for individual-level characteristics. Standard errors are clustered at the LEA district level.

\subsection{Validity of the empirical approach}

The underlying assumption of this approach is that two LEA districts with a common border are similar in all relevant characteristics except the sanction intensity. LEA districts represent relatively small geographical entities and delineations of functional local labor markets in Germany typically result in larger geographical entities \citep[see, e.g.,][who identify 50 local labor market regions, compared to 178 LEA districts]{kropp2016three}. For example, the two largest metropolitan areas in Germany -- the Rhine-Ruhr region and the Berlin-Brandenburg area -- are home to approximately 10.9 million and 6.2 million residents, respectively. At the same time, they encompass 13 and eight distinct LEA districts each. Multiple LEA districts being part of larger local labor markets makes it likely that bordering LEA districts will exhibit similar characteristics. To empirically support this premise, Table~\ref{tab:similarity} contrasts disparities in regional labor market indicators -- such as unemployment rates, vacancy rates, GDP, etc. -- within 487 pairs of neighboring LEAs with differences in randomly selected LEA district pairs \citep[see also][]{Caliendo2022}. For instance, the average disparity in unemployment rates between two randomly chosen LEA districts is approximately 4.0 percentage points. In contrast, when examining pairs of LEAs that share a common border, this disparity is markedly reduced by about 70\%, resulting in a mere 1.2 percentage point difference.

%Other regional policies (e.g., industry, infrastructure) are determined at the county, federal or state level. The regional levels of the LEA districts is only used for the administration of employment services in Germany. Therefore, it is unlikely that other types of regional policies vary at the LEA borders and affect our results.

Moreover, we conduct balancing tests regressing the local sanction intensity on a rich set of individual-level characteristics to further examine the validity of our approach. As in our main analysis, we condition on border-pair fixed effects, as well as the set of regional characteristics, and we explore the predictive power of socio-demographic characteristics, labor market histories and personality traits, all variables that have been proven to be important for individuals' labor market success. As shown in Appendix Table~\ref{tab:balancing}, we find very little evidence that individual characteristics as observed in our data are correlated with the conditional sanction intensity (i.e. see $p$-values at the bottom of Table~\ref{tab:balancing}). %Note that, in the following analysis, we condition on full set of covariates. 

%\textbf{XXX AS: Add balance tests for admin sample? XXX}

Another concern relates to the possibility that LEAs with more restrictive sanction regimes also adjust other dimensions of their policy style. In that case, any effect of the sanction intensity could possibly reflect changes in the usage of other policy instruments rather than sanctions. To test this, we exploit survey data on various dimensions of caseworkers' counseling activities including notifications about labor market programs (i.e. training, workfare programs, job creation schemes, and start-up subsidies), the number of caseworker meetings, and the provision of vacancy referrals. These variables are the most direct measures of the LEA's policy style, since they reflect the caseworkers' information strategy. The findings presented in Appendix Table~\ref{tab:sanc_counsel} provide no evidence that the local sanction risk is related to caseworkers' counseling activities. %Jason: I find this surprising no? Wouldn't I expect behavior to manifest in multiple ways? If employees of the LEA have the leniency to be more strict, wouldn't I expect them to behave dissimilarly in other regards too?

\begin{table}
\centerline{
\begin{threeparttable}
\begin{footnotesize}
\caption{Similarity of bordering regions \label{tab:similarity}}
\tabcolsep=0.45cm
\begin{tabular}{lcccc}
\hline\hline
\\[-1.3ex]
& \mc{2}{c}{Absolute difference within} &  \\
& \mc{2}{c}{pair of LEA districts} &&  \%-change \\
\\[-2.0ex]
\cline{2-3} \cline{5-5}
\\[-2.0ex]
& Simulated & Actual && Simulated - \\
& border pairs & border pairs && actual   \\
& (1) & (2) && (3) \\
\\[-1.3ex]
\hline
\\[-1.3ex]
Unemployment rate&     0.040&     0.012&&     -70.4\%\\
GDP per capita in \EUR{1,000}&     7.576&     4.948& &    -34.7\%\\
Vacancy rate &     0.066&     0.028& &    -58.6\%\\
Share of working population\\
\quad in agriculture sector &     0.017&     0.008& &    -49.6\%\\
\quad in manufacturing sector &     0.079&     0.043& &    -45.1\%\\
\quad in service sector &     0.084&     0.047&   &  -43.9\%\\
Migration rate&     0.009&     0.005&   &  -36.4\%\\
\\[-1.3ex]
\hline
\\[-1.3ex]
No. of LEA districts & 178 & 178\\ 
No. of border pairs & 1,068 & 487 &  \\
\\[-1.3ex]
\hline\hline
\end{tabular}
\end{footnotesize}
\renewcommand{\baselinestretch}{1.0}
\begin{tablenotes}\scriptsize
\item \textit{Note:} The table compares absolute differences in regional indicators within the 487 actual border pairs
to a set of simulated regional pairs. Simulated regional pairs are generated by matching each LEA district
to three other randomly selected LEA districts, which yields 1,068 simulated pairs. Columns (1) and (2) report absolute average
differences in regional indicators across all actual and simulated pairs, respectively. Column (3) reports the \%-difference
between average differences reported in columns (2) and (1). 
\end{tablenotes}
\end{threeparttable}}
\end{table}

\newgeometry{top=20mm,bottom=20mm}
\begin{table}[p!]
	\centerline{
		\begin{threeparttable}
			\begin{footnotesize}
				\caption{Balancing test of sanction intensity (survey data) \label{tab:balancing}}
				\tabcolsep=0.5cm
				\begin{tabular}{lcc}
					\hline\hline
					\\[-2.0ex]
Dependent variable					& \mc{2}{c}{Sanction intensity $TI_j$}  \\
					\\[-2.0ex]
					\cline{2-3}
					\\[-1.8ex]
					& Coef. & SE  \\
					\hline
					\\[-1.8ex]
\textbf{Education}\\
School leaving degree (Ref.: None)&        ref.        &                      \\
\quad Lower sec. degree         &         -0.239        &     (0.197)\\
\quad Middle sec. degree        &         -0.091        &     (0.170)\\
\quad Upper sec. degree        &         -0.069        &     (0.172)\\
Higher education (Ref.: None)&        ref.        &                   \\
\quad Vocational training   &          0.243        &     (0.156)\\
\quad University degree  &          0.264        &     (0.178)\\
\mc{3}{l}{\textbf{Socio-demographic characteristics}}\\
Female              &            0.010        &     (0.050)\\
Migration background           &          -0.031        &     (0.092)\\
Age (Ref.: 16-24 years)&             ref.        &            \\
\quad 25-34 years     &            0.065        &     (0.074)\\
\quad 35-44 years      &         0.035        &     (0.086)\\
\quad 45-55 years      &          -0.023        &     (0.088)\\
Married             &         -0.071$^{*}$  &     (0.042)\\
Children (Ref.: None)&               ref.        &            \\
\quad One child           &          0.029        &     (0.065)\\
\quad Two or more children           &          0.087        &     (0.096)\\
Homeowner           &          0.103        &     (0.064)\\
\mc{3}{l}{\textbf{Unemployment and labor market history}}\\
UI Benefit recipient            &        -0.001        &     (0.071)\\
Level of UI benefits in \euro 100 & -0.005 & (0.006)\\
Lifetime months in unemployment (div. by age-18) &        -0.228 &     (0.210)\\
Lifetime months in employment (div. by age-18) &         -0.100        &     (0.091)\\
Last hourly wage in \EUR{}    &         0.012$^{*}$  &     (0.006)\\
Employment status before UE (Ref.: Other)&        ref.        &       \\
\quad Regular employed         &          -0.020        &     (0.103)\\
\quad Subsidized employed          &         -0.064        &     (0.111)\\
\quad School, apprentice, military, etc.         &       -0.023        &     (0.126)\\
\quad Parental leave         &       -0.269$^{*}$  &     (0.143)\\
\mc{3}{l}{Time between entry into UE and interview (Ref.: 7 weeks)}\\
\quad 8 weeks      &         -0.052        &     (0.248)\\
\quad 9 weeks       &         -0.076        &     (0.278)\\
\quad 10 weeks     &         0.046        &     (0.274)\\
\quad 11 weeks      &         -0.005        &     (0.313)\\
\quad 12 weeks      &         -0.104        &     (0.316)\\
\quad 13 weeks     &        -0.023        &     (0.344)\\
\quad 14 weeks      &       -0.066        &     (0.344)\\
\textbf{Personality traits}$^{(a)}$\\
Openness     &         -0.019        &     (0.021)\\
Conscientiousness     &         -0.023        &     (0.025)\\
Extraversion      &            0.044        &     (0.040)\\
Neuroticism     &           0.023        &     (0.029)\\
Locus of control     &         -0.040        &     (0.036)\\
Constant            &         -0.058        &     (0.472)\\
\hline
%Observations        &       31857        &            &       31857        &            &       31857        &            \\
No. of observations     &    5,669        &                   \\
Additional control variables\\
\quad Federal state fixed effects & Yes & \\
\quad Basic regional characteristics & Yes & \\
$P$-value joint significance\\
\quad Education       &     0.454        &            \\
\quad Socio-demographic characteristics      &     0.238        &            \\
\quad Unemployment and labor market history   &      0.286        &            \\
\quad Personality traits      &      0.526        &            \\
\\[-1.8ex]
\hline\hline
\end{tabular}
\end{footnotesize}
\begin{tablenotes}\scriptsize
\item \textit{Note:} The table reports the results of balancing tests regressing the local treatment intensity on individual-level characteristics. Standard errors in parenthesis are clustered at the LEA district level. $^{***}/^{**}/^{*}$ indicate statistical significance at the 1\%/5\%/10\%-level.
\item $^{(a)}$Personality traits are measured with different items on a 7-point Likert-scale and standardized to have a mean of zero and a variance of one.
\end{tablenotes}
\end{threeparttable}}
\end{table}
\restoregeometry

\begin{table}[h!]
	\centerline{
		\begin{threeparttable}
			\caption{Sanction intensity and counseling activities of caseworkers}\label{tab:sanc_counsel}
			\tabcolsep=0.35cm
			\begin{footnotesize}
				\begin{tabular}{lcccccc}
					\hline\hline 
					\\[-2.0ex]
					& \mc{4}{c}{Notification about labor market programs} & & \\
							\\[-2.0ex]
					\cline{2-5}
							\\[-2.0ex]
					&   &  Job   &  &  & Any  & Caseworker\\ 		
Dependent variable	&  Workfare &  creation & Training & Start-up &  vacancy &  meetings:  \\
				& program	&  schemes		 &	program  & subsidies & referral & three or more  \\
						\\[-2.0ex]
					& (1) & (2) & (3) & (4) & (5) & (6) \\
	\\[-2.0ex]
\hline
\\[-2.0ex]					
Sanction intensity   &      0.004 &      -0.054 &      0.125 &      -0.109 &      -0.083 &  0.244 \\
         &      \raisebox{.7ex}[0pt]{\scriptsize(0.402)} &      \raisebox{.7ex}[0pt]{\scriptsize(0.021)} &      \raisebox{.7ex}[0pt]{\scriptsize(0.114)} &      \raisebox{.7ex}[0pt]{\scriptsize(0.078)} &      \raisebox{.7ex}[0pt]{\scriptsize(0.105)} & \raisebox{.7ex}[0pt]{\scriptsize(0.164)}\\
 %        	\\[-2.0ex]
%         \hline
         \\[-2.0ex]	
No. of observations         &   5,669 &  5,669 &   5,669 &  5,669 &  5,669&  5,669 \\
Mean dep. variable   &      0.014&      0.020 &      0.164 &      0.048&     0.239& 0.634\\
\\[-2.0ex]
\hline \hline
\end{tabular}
\end{footnotesize}
\begin{tablenotes} \scriptsize 
\item \textit{Note:} The table reports the effect of the local sanction intensity on caseworkers' counseling activities. In all specifications, we account for socio-demographic and regional characteristics, as well as border-pair fixed effects. Standard errors clustered at the LEA district level are shown in parentheses. ***/**/* indicate statistical significance at the 1\%/5\%/10\%-level.
\end{tablenotes}
\end{threeparttable}}
\end{table}

\end{appendix}

\end{document}